\documentclass[pre,aps,floatfix,showpacs,showkeys]{revtex4}

\usepackage{amssymb}
\usepackage[centertags]{amsmath}
\usepackage{latexsym}
\usepackage[dvips]{graphicx}
\usepackage{epsfig}
\usepackage{pstricks}
\usepackage{subfigure}
\usepackage[]{hyperref}

\begin{document}
\author{C. Fortmann, T. D\"oppner,
A. Przystawik, T. Raitza, G. R\"opke, J. Tiggesb\"aumker, N.X. Truong}
\affiliation{Universit\"at Rostock, Institut f\"ur Physik,
18051 Rostock, Germany}
\author{T. Laarmann}
\affiliation{Max-Born-Institut, Max Born Strasse 2a, 12489 Berlin,
Germany}
\title{On the occurrence of Balmer spectra in expanding microplasmas
from laser irradiated 
liquid hydrogen}
\date{\today}
\begin{abstract}
	\noindent
        Balmer spectra are investigated which are obtained from 
        hydrogen droplets irradiated by ultra-short intense laser
        pulses. A unified quantum statistical 
	description of bremsstrahlung, the Stark broadening
        and the van der Waals profile of hydrogen spectral lines is
        used, which allows to include many-particle
        effects.  Analyzing the line profiles, a low ionization degree
        of a dense plasma is inferred, where the main contribution to
        the spectral line shape originates from the interaction with
        the neutral components. Effective temperatures and densities
	of the radiating microplasma are deduced.
	A dynamical description is given within plasma hydrodynamics, 
	explaining
        the formation of excited atomic states in the expanding
        system and the occurrence of the observed Balmer lines
	only below a critical density. 

\end{abstract}

\maketitle

\section{Introduction\label{sec:intro}}
Spectroscopy is one of the most powerful methods in plasma
diagnostics. In particular, the analysis of the shape of spectral
lines allows to determine the properties of the plasma such as
temperature, density, and composition. For dense, strongly coupled
plasmas, optical spectra have been investigated to infer the parameter
values not only of laboratory, but also of astrophysical plasmas,
see, e.g. Refs.
\cite{Flohr-Piel_PRL70_1108_1993,Godbert_PRE49_5889_1994,
  Malyshev_PRE60_6016_1999,Tadokoro_PRE57_R43_1998} 
and Refs.  \cite{Soria_APJ539_2000,Benz_plasma_astrophys,rybi},
respectively.

The shape of spectral lines is determined by different processes.
Besides the natural line width given by the finite lifetime of excited
states, Doppler broadening is related to the thermal motion of the
emitters. The influence of the surrounding medium becomes more
important at increasing density and leads to pressure broadening. A
large effect is caused by charged particles in the plasma. This Stark
broadening is determined by the distribution of ions as sources of the
microfield, whereas the contribution of the electrons is usually 
treated in impact approximation
\cite{Baranger,GriemKolb}. The Weisskopf radius can be used to
distinguish between weak and strong collisions. 
However, also the neutral components in the surrounding
medium contribute to line profile, i.e. as
van der Waals broadening. 

A systematic approach to the shape of spectral lines in dense plasmas
has been worked out on the basis of perturbation theory
\cite{Sobelmann}, and
many-particle effects have been incorporated. 
Griem \cite{griem_plasma_spectroscopy} gives a review of different
classical and semiclassical methods as well as 
their application to line shapes in different plasmas. The
\emph{unified} theory \cite{Smith_PhysRev1969,Voslamber_ZsNtf1969} was
developed to describe the center as well as the wings of the lines.
A quantum statistical
approach based on the many-particle Green functions method was
presented by G\"unter \cite{Guenter_PRA44_1991,guen:habil95} and has
been developed systematically for charged perturbers.
Special attention was payed to show the equivalence between this
rigorous approach and the aforementioned classical and semiclassical
concepts in the appropriate limits.  The quantum statistical
approach has been applied to a broad range of systems, e.g. hydrogen
plasmas
\cite{Roepke_AnPh38_1981,Hitzschke_JPhysB19_2443_1986,
  Guenter_PRA44_1991,Koenies_JQSRT52_423_1994}, 
helium plasmas \cite{Milosavljevic_EPJD23_385_2003,Omar_PRE73_2006}
and plasmas of other elements containing H or He like ions.  In most
cases, a hot and dense plasma with a large fraction
of free electrons was considered. In such systems, the main
contribution to the shift and width of a given spectral line, besides
the natural line width and Doppler broadening, comes from the ionic
microfield, caused by ions surrounding the radiating atom or ion,
collisions with free electrons, and interaction with collective
excitations in the plasma (plasmons). Theoretical results are in very
good agreement with experiments carried out, e.g. by Wilhein
\cite{Wilhein_JOptSocAmB15_1235_1998,Sorge_JPhysB33_2983_2000}.

In weakly ionized plasmas, the influence of neutral perturbers becomes
dominant. Van der Waals
broadening has been investigated within second order perturbation
theory. Based on the derivation of the van der Waals potential between
two hydrogen atoms in their ground state as given by London
\cite{EisenschitzLondon_ZsfPh60_491_1930}, a red shift of the spectral
line follows as was shown by Margenau \cite{Margenau_prev48_755_1935}.
A review of van der Waals broadening is given by Traving
\cite{traving_in_LH}. Line broadening due to van der Waals interaction
has been investigated in detail, e.g. by Walkup \emph{et al.}
\cite{Walkup_PRA29_169_1984}.

We show how to extend the quantum statistical approach, elaborated for
the influence of charged particles on the line profile, to include also
the interaction with neutral perturbers. In this way, the extension of
the van der Waals broadening by accounting for many-particle effects
such as dynamical screening or strong collisions is possible. A
unified description of the influence of the perturbing partially
ionized plasma on the line profile of a radiating atom can be
performed, treating charged particles and neutrals within the same
formalism.

The theory will be compared to optical spectra
obtained from hydrogen microdroplets exposed to strong femtosecond
laser pulses. So far, most of the experimental and theoretical
work has concentrated on nanosized clusters
\cite{DitPRA96,DoePRL05,KimPRL03}.
Studies on the light emission from these systems have focused on
x--rays and EUV--radiation \cite{ParPRE00,SchEPJD01,DueAPB01},
motivated, e.g. by the search for novel light sources. Hydrogen as a
target material has attracted considerable attention since nuclear
fusion has been demonstrated as a result of Coulomb explosion
of dense deuterium clusters \cite{MadPRA04}. Various efforts to produce
micrometer sized hydrogen targets are reported, cf. e.g.
Ref.~\cite{NorNIMPRA05}. Other studies on such large systems
include, e.g. the  measurement of x--ray emission from Kr
clusters \cite{HanPRE05} and from methanol microdroplets
\cite{Anand_ApPhb81_469_2005,AnaOE06}.
An expanding microplasma can be
produced with interesting parameter values. In particular, strongly
coupled plasmas are obtained, where density and temperature are
time-dependant. Time resolved experimental techniques
have been demonstrated to measure transient plasma
properties like the evolution of the plasma
density \cite{ZweOE00,LiuPRA06}.

In the experiment on $\mu$m sized droplets reported here, 
lines of the hydrogen Balmer series are observed as distinctive
features in the optical emission spectrum. 
The present paper will give a first
interpretation of the recorded line shapes. 
The analysis of the line profiles shows that van der Waals 
broadening due to neutral perturbers is the predominant contribution. 
Values for the effective plasma parameters are deduced to interprete
the observed line profiles, which differ for the different lines. However, an
equilibrium picture cannot give an agreeable description of the
measured Balmer spectra. A consistent dynamical picture of the expanding
microplasma will be given by use of hydrodynamic simulations.
This enables us
to interpret the observed spectra if considering
the time stages where excited atomic states can exist.

The work is organized as follows: In the second section we will
briefly outline the method of thermodynamic Green functions with
emphasis on its application to the calculation
of spectral line shapes.  
The method will be applied to the case of a weakly ionized plasma, where
the interaction among atoms is governed by the dipole-dipole term. The
Margenau result 
appears in the case of ground state perturbers.
In the third section, our results will be compared to the Balmer spectra
obtained from laser produced hydrogen microplasmas. Parameter values
for the effective temperature and density are inferred.  In the fourth section, a hydrodynamical
description of the expanding droplet after laser excitation is given.

\section{Many-body theory of line shapes and van der Waals
  broadening\label{sec:theory_lineshapes}}
\subsection{Quantum statistical approach 
to spectral line shapes\label{subsec:quantstat}}

A quantum statistical approach to the optical spectra of dense,
strongly coupled systems can be given within linear response theory.
Emission and absorption of radiation is related to the transverse
dielectric function $\epsilon_{\rm tr}(\vec k,\omega)$. In the optical
region, the wavelength is large compared with atomic distances so that
the long-wavelength limit $k \to 0$ can be considered. For the
absorption coefficient we find
\begin{equation}
	\alpha(\omega) =\frac{\omega}{cn(\omega)}\mathrm{Im}\,
        \epsilon(0,\omega)~, 
\end{equation}
$n(\omega)$ is the refraction index, $c$ is the velocity of light in
vacuum. For $k \to 0$ the transverse and longitudinal dielectric
function $\epsilon(0,\omega)$ become identical. We consider in the
following the longitudinal one.

The dielectric function for a charged particle system can be evaluated
in a rigorous way using quantum statistical methods.  According to
$\epsilon(\vec k,\omega)=1-(1/\epsilon_0 k^2) \Pi(\vec k,\omega)$, it
is linked to the polarization function $\Pi(\vec k,\omega)$, for which
a systematic perturbation expansion can be given. A Green function
approach to the polarization function is outlined in
App.~\ref{app:pi_2}. Using Feynman diagrams and partial summations,
appropriate approximations can be found which account for different
microscopic processes contributing to the behavior of the charged
particle system. In particular, for partially ionized plasmas,
a
cluster decomposition $\Pi(\vec k,\omega)=\Pi_1(\vec
k,\omega)+\Pi_2(\vec k,\omega)+\ldots$ can be performed to give the
contribution of free carriers as well as of bound states in a
systematic way \cite{RoepkeDer_physstatsolb92_501_1979}.
 
$\Pi_1(\vec k,\omega)$ collects the single-particle contribution. In
lowest order, neglecting collisions with other particles, the
well-known RPA polarization function
\begin{equation}
\Pi_1^{(0)}(\vec k,\omega)=\Pi^\mathrm{RPA}(\vec k,\omega)=\sum_{c,\vec
  p}e_c^2 { f_c(\vec
p)-f_c(\vec p-\vec k) \over E^c_p-E^c_{\vec p-\vec
k}-\hbar \omega-i\eta }
\end{equation}
is obtained. Here, $f_c(\vec p)=[\exp\{(E^c_p-\mu_c)/k_BT\}+1]^{-1}$ is
the Fermi distribution function of particles of species $c$ (including
spin) with
charge $e_c$ and mass $m_c$, $E^c_p = \hbar^2 p^2/2 m_c$ is the
single-particle kinetic energy, and $\mu_c$ the chemical potential.
Within this lowest order approximation, single-particle excitations as
well as collective plasmon modes are described. The limit $\eta\to
0^+$ has to be performed after the summation over momenta.

The RPA contribution $\Pi_1^{(0)}(\vec k,\omega)$ is improved if
interactions with other particles beyond mean field are included, see
App.~\ref{app:pi_2}. Medium effects, such as collisions, dynamical
screening and exchange, enter the polarization function via the
single-particle self-energies and the vertex function. The vertex
function describes the in-medium coupling of particles to the
radiation field.  It has to be taken in the same approximation as the
self-energy as dictated by consistency constraints, i.e. Ward
identities \cite{maha}. Going beyond the mean-field (Hartree-Fock)
approximation, collisions can be considered in lowest order Born
approximation. Higher orders lead to dynamical screening and t-matrix
expressions for strong collisions. This way, the continuum of optical
spectra is described, in particular inverse bremsstrahlung
\cite{wier:phpl01}. However, line spectra are missing in
$\Pi_1(\vec k,\omega)$.

The next term in the cluster decomposition is $\Pi_2(\vec k,\omega)$.
It is given by the convolution of two atomic (two-particle)
propagators and describes also transitions between bound states, i.e.
the line spectrum \cite{RoepkeDer_physstatsolb92_501_1979}. In the
lowest order, any interaction with
further particles are neglected. Similar to $\Pi^{(0)}_1(\vec
k,\omega)$ given above, replacing single-particle propagators by
atomic propagators one obtains
\begin{equation}
	\Pi^{(0)}_2(\vec k,\omega)=4\sum_{\alpha_1\alpha_2,\vec
	P}\left|M_{\alpha_1\alpha_2}^{(0)}(\vec
	k)\right|^2
	{g(E^{(0)}_{\alpha_1\vec P})-g(E^{(0)}_{\alpha_2\vec P-\vec k})\over
	E^{(0)}_{\alpha_1\vec P}-E^{(0)}_{\alpha_2\vec P-\vec k}- \hbar
        \omega-i\eta }~. 
	\label{eqn:Pi_20_sigma}
\end{equation}
The factor 4 accounts for spin degeneration,
$g(E^{(0)}_{\alpha \vec P})=[\exp\,\{(E^{(0)}_{\alpha \vec
  P}-\mu_\mathrm{i}-\mu_\mathrm{e})/k_BT\}-1]^{-1}$ is  
the Bose distribution function of electron-ion bound states at
$E^{(0)}_{\alpha P}=-Z^2\,\mathrm{Ry}/n^2+ \hbar^2 P^2/2M$, 
with $\alpha=\{n,l,m,m_s\}$ 
denoting the internal quantum numbers, $\hbar \vec P$ the center of mass
momentum, and $M=m_\mathrm{i}+m_\mathrm{e}$.
The unperturbed matrix elements $M_{\alpha_1\alpha_2}^{(0)}(\vec
	k)$ are given by
\begin{align}
	\label{eqn:matrixelementM_p}
	M_{\alpha_1\alpha_2}^{(0)}(\vec
	k)&=e\sum_p\psi_{\alpha_1}^*(\vec
	p)\left[\psi_{\alpha_2}(\vec p-\frac{m_\mathrm{e}}{M}\vec
	k)-\psi_{\alpha_2}(\vec p+\frac{m_\mathrm{i}}{M}\vec k)  \right]\\
	&\simeq e\left[ \delta_{\alpha_1\alpha_2}- \int\!\mathrm{d}^3\vec
	r\,\psi_{\alpha_1}^*(\vec r)\,\mathrm{e}^{i\vec k\cdot\vec
	r}\,\psi_{\alpha_2}(\vec r)\right]~.
	\label{eqn:matrixelementM_r}
\end{align}
$\psi_\alpha(\vec p)=\langle\,\vec p\,|\,\alpha\,\rangle$ and
$\psi_\alpha(\vec
r)=\langle\,\vec r\,|\,\alpha\,\rangle$ are atomic wave-functions in
relative momentum representation and coordinate
representation, respectively. 
Eq.~(\ref{eqn:matrixelementM_r}) results for $m_\mathrm{e}/m_\mathrm{i}\to 0$.
In contrast to a simple chemical picture where a
mixture of free charge carriers and atoms is considered, scattering
states are included, and double counting of diagrams has to be avoided.
In this approximation, one obtains the unperturbed line-spectrum of
isolated atoms including the Doppler profile due to thermal motion.

As before, medium effects are described by taking the self-energy of
the constituents and the vertex into account, which now are given on
the two-particle level, see App.~\ref{app:pi_2}. In the simplest
approximation, the dynamically screened interaction can be considered
similar to the $GW$ approximation for the single-particle propagator
\cite{Fortmann_CPP_accepted_2006}.
This dynamically screened interaction contains the polarization
function, which once more can be decomposed into the contribution of
single-particle states, two-particle states and higher cluster states.
In this way, we obtain the dynamically screened Born approximation for
collisions with free particles or composed clusters of the partially
ionized plasma. 

In the case of dense and strongly ionized systems, the main
contribution to the self-energy comes from the interaction of the
radiating atom with charged particles, i.e. electrons and ions.  The
influence of free particles is given by the RPA-polarization function,
i.e. the first term in the cluster decomposition. This interaction can
be inserted into the self-energy, but the vertex-function has to be
taken in the same approximation. Reducing the screened interaction to
its one-loop approximation, which amounts to the second Born
approximation with respect to the statically screened Coulomb
potential, the so-called impact approximation is obtained.
Systematic improvements by taking higher order contributions into
account lead to the dynamical screening of the Coulomb potential in
the case of weak collisions and the t-matrix approach in the case of
strong collisions. In particular, the contribution of collisions with
electrons to the line profile in dense plasmas is treated this way,
for a review see G\"unter \cite{guen:habil95}.

The following expression for the line-profile $\mathcal
L(\Delta\omega)$ as a function of the frequency displacement
$\Delta\omega=\omega-(E_{i}^{0}-E_{f}^0)/ \hbar$ is obtained
\cite{guen:habil95},  
\begin{multline}
	\mathcal L(\Delta\omega)=\sum_{ii'ff'}M_{if}^{(0)}(\vec k)\left[
	M_{i'f'}^{(0)}(\vec k) \right]^*
	\frac{\mathrm{e}^{-\beta
            \hbar(\omega_{if}+\Delta\omega)}\left(\omega_{if} 
	+\Delta\omega \right)^4}{8\pi^3c^3}
	\int\frac{\mathrm{d}^3 P}{(2\pi)^3}\,\mathrm{e}^{-\beta\frac{
	\hbar^2
        P^2}{2M}}\,\int_0^\infty\mathrm{d}\bar\beta\,W_\rho(\bar\beta) 
	\\
	\times
	\langle\left. \,i\,\right|\langle\left. \,f\,\right|\left[
	\hbar \Delta\omega-\frac{\hbar^2 \vec P\cdot\vec
	k}{M}-\frac{\hbar^2 k^2}{2M}-\mathrm{Re}\,\left\{
	\Sigma_i(\Delta\omega,\bar\beta)-\Sigma_f
	\right\}+i\mathrm{Im}\,\left\{
	\Sigma_i(\Delta\omega,\bar\beta)+\Sigma_f \right\}+i\Gamma^v
	\right]^{-1}\left|\,f'\,\right.\rangle\left|\,i'\,\right.\rangle~.
	\label{eqn:lineprofile_sigma}
\end{multline}
Here, $\Sigma_i,\Sigma_f$ denote the electronic contribution to the
self-energy of the initial or final state, respectively, $\Gamma^v$
the vertex contribution. Improvements of the Born approximation
by accounting for dynamical screening and strong
collisions, have extensively been discussed in the literature
\cite{griem_plasma_spectroscopy}.
 
Due to their large masses, ions are much slower than electrons.  In
the adiabatic limit, the ions may be considered as a static
distribution of charged particles during the process of emission.
Considering different ionic configurations in the plasma, the
distribution $W_\rho(\bar\beta)$ of the so-called microfield
$\bar\beta =E/E_{\rm Holtsmark}$, $E_{\rm Holtsmark}= (4
n_\mathrm{ion}/15)^{2/3}  Ze/(2
\epsilon_0)$, ($n_\mathrm{ion}$ is the density of ions with charge Z in the system) is
introduced, and the averaging leads to the Stark profile as given in
Eq. (\ref{eqn:lineprofile_sigma}). Also in the case of ionic
contributions, the systematic many-particle treatment leads to further
improvements such as the dynamic microfield.  

In the next step, one has to take into account the interaction between the
radiator and neutral particles (atoms), as well. This can be performed
along the same lines as for the interaction with charged particles,
discussed before.
The bound state contribution is considered in the
cluster decomposition of the polarization function, expanding the
dynamically screened interaction which enters the self-energy, see App.~\ref{app:pi_2}. In particular we
will consider the two-particle self-energy describing the influence of
the medium on the radiator, where the bound state component is taken
into account in lowest approximation. We consider the second order
Born approximation with respect to the unscreened Coulomb potential to
describe the interaction between neutrals. The detailed calculation is
given in the next section.

Again, improvements can be obtained systematically. By
calculating the dynamically screened interaction with
 the polarization function up to the second term 
in the cluster decomposition, the interaction with free and bound
particles as well as with collective excitations could be included in
a consistent way.
Strong collisions can be described via the four-particle t-matrix. In
addition to the self-energy term, also the vertex is modified if bound
states are included.
 
\subsection{Evaluation of atom-atom interaction in Born approximation\label{sec:bound-bound_born}}
The calculation of the interaction $V_{\alpha_1\alpha_2}(R)$ between
two atoms at 
distance $R$ in state
$\left|\alpha_1\!\right.\rangle$ and $\left|\alpha_2\!\right.\rangle$
is carried out in
App.~\ref{app:bound}. We obtain 
\begin{eqnarray}
	V_{\alpha_1\alpha_2}(R)&=& {e^4 \over (4 \pi \epsilon_0)^2}
        \sum_{\alpha_3\alpha_4} \frac{1}{
	E^{(0)}_{\alpha_1}+E^{(0)}_{\alpha_2}-E^{(0)}_{\alpha_3}-E^{(0)}_{\alpha_4}}\nonumber\\
	&&\times
	\left|\sum_{\vec r_1\vec r_2}
	\psi_{\alpha_1}^*(\vec r_1)\psi_{\alpha_3}(\vec r_1)
	\left[ \frac{1}{R}-\frac{1}{|\vec R-\vec r_1|}-
	\frac{1}{| \vec R+\vec r_2 |}
	+\frac{1}{|\vec R-\vec r_1+\vec r_2|}\right]
	\psi_{\alpha_2}^*(\vec r_2)\psi_{\alpha_4}(\vec r_2)
	\right|^2~.
	\label{eqn:V_vdw_R}
\end{eqnarray}
Here, $\alpha_3, \alpha_4$ denote quantum numbers of the intermediate
states. Only elastic processes
are considered where the internal quantum numbers for the initial
state and the final state are identical. Due to the ion's
heavy mass, we neglect the contribution of kinetic energy in the
intermediate propagator in comparison with the excitation energies.
Besides $R$ which is the distance between the nuclei of the
interacting atoms, $\vec r_1$ and $\vec r_2$ are vectors of position
of the two electrons of atom 1 and atom 2, respectively.

This result (\ref{eqn:V_vdw_R}) coincides with results from second
order perturbation theory.
At large distances $R/a_B \gg 1$ we can perform the dipole
approximation using
$|\vec R-\vec r|^{-1} \approx 1/R^2\left[R-(R^2+r^2-2\vec R\cdot\vec
r)/2R+3(R^2+r^2-2\vec R\cdot\vec r)^2/8R^2\right]$ for $R\gg r$. We
obtain the van der Waals potential,
\begin{eqnarray}
	V^{\rm vdW}_{\alpha_1\alpha_2}(R)&=&\frac{e^4}{ (4 \pi
          \epsilon_0)^2 R^6}
		\sum_{\alpha_1\alpha_2}\frac{1}{
		E^{(0)}_{\alpha_1}+E^{(0)}_{\alpha_2}-E^{(0)}_{\alpha_3}-E^{(0)}_{\alpha_4}}\nonumber\\
		&&\times
		\left|\sum_{\vec r_1\vec r_2}
		\psi_{\alpha_1}^*(\vec r_1)\psi_{\alpha_3}(\vec r_1)
		\left[\vec r_1\cdot \vec r_2-3\frac{\left(\vec r_1\cdot\vec
		R\right)\;\left(\vec r_2\cdot\vec R\right)}{R^2}\right]
		\psi_{\alpha_2}^*(\vec r_2)\psi_{\alpha_4}(\vec r_2)
		\right|^2~.
	\label{eqn:V_vdw_R_dip}
\end{eqnarray}
For hydrogen-like atoms, the sum in 
Eq.~(\ref{eqn:V_vdw_R_dip}) has been performed by Eisenschitz and London
\cite{EisenschitzLondon_ZsfPh60_491_1930}. Their result for the
interatomic potential between two hydrogen atoms in ground state
($\alpha_1=\alpha_2=1s$) is given by 
\begin{equation}
	V^\mathrm{vdW}_{1s,1s}(R)=
	-\frac{C_6}{R^6}~.
	\label{eqn:vdwpot_00}
\end{equation}
The constant
$C_6$ was calculated as $C_6=12.94 \,\mathrm{Ry}\,\,
a_\mathrm{B}^6$, where 
$\,\mathrm{Ry}=13.6\,\mathrm{eV}$ is the Rydberg energy and 
$a_\mathrm{B}=0.53$\,\AA\ is the Bohr radius.

For the interaction between a radiator in an excited state $\alpha =
n,l$ and a perturber in ground state, $\alpha=1s$, the following
expression for the 
interaction strength $C_6(\alpha,1s)$, can be derived, as shown in
App.~\ref{app:interaction_strength},
\begin{equation}
	C_6(\alpha,1s)=4 \frac{n^2}{1+1/n^2}\left[ 5n^2+1-3l(l+1) \right]~.
	\label{eqn:int_strength}
\end{equation}
 
Note that there is no exchange contribution in
Eq.~(\ref{eqn:V_vdw_R}), since the distance $R$ is assumed to be large
compared with the Bohr radius. At shorter distances, the higher order
terms as well as exchange terms have to be taken into account which
will diminuish the van der Waals behavior at small distances and
possibly, in the case of electrons in the triplet state, result in a
repulsion due to the exchange contribution.


\subsection{Van der Waals profiles}

Within a general approach, the contribution of the dense medium to the
spectral line shape is given by the self-energy and vertex
contribution. The first order of the interaction gives the
Hartree-Fock mean field, in second order the impact
approximation is obtained
where collisions are described in Born approximation.

Within this concept, the van der Waals contribution to the self-energy
is
\begin{equation}
	\Sigma_{\alpha_1}^{\rm vdW} = \sum_{\alpha_2,P}
	g(E^{(0)}_{\alpha_2\vec P}) V_{\alpha_1\alpha_2}^{\rm
	vdW}(q=0)~,
\end{equation}
which in coordinate space reads 
\begin{equation}
	\Sigma_{\alpha_1}^{\rm vdW} = \sum_{\alpha_2}n_{\alpha_2} \int d^3R\,\, V_{\alpha_1\alpha_2}^{\rm vdW}(R)~.
\end{equation}
Here, $n_{\alpha_2}$ is the density of pertubing atoms in state
$\alpha_2$.
In the following, we consider the
perturbing atoms in their ground state $\alpha_2\equiv 0$
\cite{note_groundstate}
and only the
radiator in an excited state $\alpha=nl,\,n>1$.

As already discussed in connection with the ionic contribution to the
spectral line shapes, because of their large masses the atomic 
motion is slow,
and we can apply the microfield concept of strong interaction with a
given field distribution.
The intensity distribution of a given spectral line, due interaction
between the radiator in initial and final states
$\left|i\!\right.\rangle$ and $\left|\,f\,\right.\rangle$,
respectively and the perturber in 
its ground state $\left|0\!\right.\rangle$ 
is obtained by
performing an averaging
procedure over the distribution of perturbing atoms,
\begin{equation}
	I_{if}(\Delta\omega)=
	\int\mathrm{d}^3\vec r_1\int\mathrm{d}^3\vec
	r_2\ldots\int\mathrm{d}\vec r_N P_N(\vec r_1,\vec
	r_2,\ldots,\vec r_N)\,\delta\left( \sum_{j=1}^N
	V_{i0}(r_j)-V_{f0}(r_j)-\hbar \Delta\omega
      \right)~. 
	\label{eqn:profile_at-at}
\end{equation}
Here, $P_N(\vec r_1,\ldots\vec r_N)$ is the probability density for
having atom 1 in the volume element $\mathrm{d}^3\vec r_1$ at $\vec
r_1$, atom 2 in the volume element $\mathrm{d}^3\vec r_2$ at $\vec
r_2$, etc. 
 The evaluation of the sum (\ref{eqn:profile_at-at}) in the case of
 statistically independent atoms, i.e. $P_N(\vec r_1,\ldots\vec
 r_N)=1/V^N$, is shown in
 App.~\ref{app:vdw-profile}.
 
The result for the dipole limit of the interaction
Eq.~(\ref{eqn:vdwpot_00}) is \cite{Margenau_prev48_755_1935},
\begin{equation}
	I^{\rm vdW}_{if}(\Delta\omega)=\Lambda_{if}\left[ -\Delta\omega
\right]^{-3/2}\mathrm{e}^{\pi\Lambda_{if}^2/\Delta\omega}~,
	\label{eqn:profile_london}
\end{equation}
with $\Lambda_{if}=(2\pi/3)\,(C_6(i,0)-C_6(f,0))^{1/2}n_\mathrm{atom}$. 
This is known as the Margenau profile.
Note that $\Delta\omega$ is to be
taken at negative values, i.e. the van der Waals interaction shifts
the spectral line to smaller energies (red shift). 
The maximum of the intensity distribution (\ref{eqn:profile_london})
is located at
\begin{equation}
	\Delta\omega_\mathrm{max}=-\left( \frac{2\pi}{3}
	\right)^{3}C_{6}\,\left( n_\mathrm{atom}\,\,a_\mathrm{B}^3 \right)^2
        \,\,\mathrm{Ry}/\hbar~,
	\label{eqn:shift_max}
\end{equation}
with $C_6=C_6(i,0)-C_6(f,0)$ and $n_\mathrm{atom}$ the number density of (ground
state) hydrogen atoms in the system.  The full width of half maximum
(FWHM) of the line is found as
$\gamma=2.78\,\Delta\omega_\mathrm{max}$, see also Refs.
\cite{Margenau_prev48_755_1935,lochte-holtgreven_in_LH}.
The Margenau profile Eq.~(\ref{eqn:V_vdw_R_dip}) will be used in order
to analyse the hydrogen Balmer spectra obtained from laser produced
microplasmas. These experiments will be described in the following
section.
\section{Experiments with laser-irradiated hydrogen
droplets \label{sec:experiments}}
\subsection{Experimental setup \label{subsec:setup}}
 Hydrogen droplets
are produced by expansion of pre-cooled H$_2$ gas at temperatures
between 17 and 22\,K through a 20\,$\mu$m orifice into a vacuum
chamber using a backing pressure of about 8\,bar. At these source
conditions, H$_2$ droplets of about $10\,\mu$m in diameter are emitted
from the nozzle at a density of $n_\mathrm{liq}=4.2\cdot
10^{22}\,\mathrm{cm}^{-3}$. Roughly 10\,mm behind the nozzle the
droplets are irradiated by intense IR laser pulses generated by a
Ti-Sapphire
laser system with a wavelength of $\lambda_0=810\,\mathrm{nm}$ and
a pulse energy of $E_\mathrm{pulse}=2.5\,\mathrm{mJ}$. The beam is
focussed to a spot with a beam waist of $40\,\mu\mathrm{m}$. The
pulse width $\tau$ is tuned in the range of
$\tau=50-1000\,\mathrm{fs}$, corresponding to laser peak
intensities between
$7.0\times10^{15}\,\mathrm{W}\,\mathrm{cm}^{-2}$ and $3.5\times
10^{14}\,\mathrm{W}\,\mathrm{cm}^{-2}$.
When the droplets are exposed to the strong laser field, optical
emission from the hydrogen microplasmas is observed.
\begin{figure}[ht]
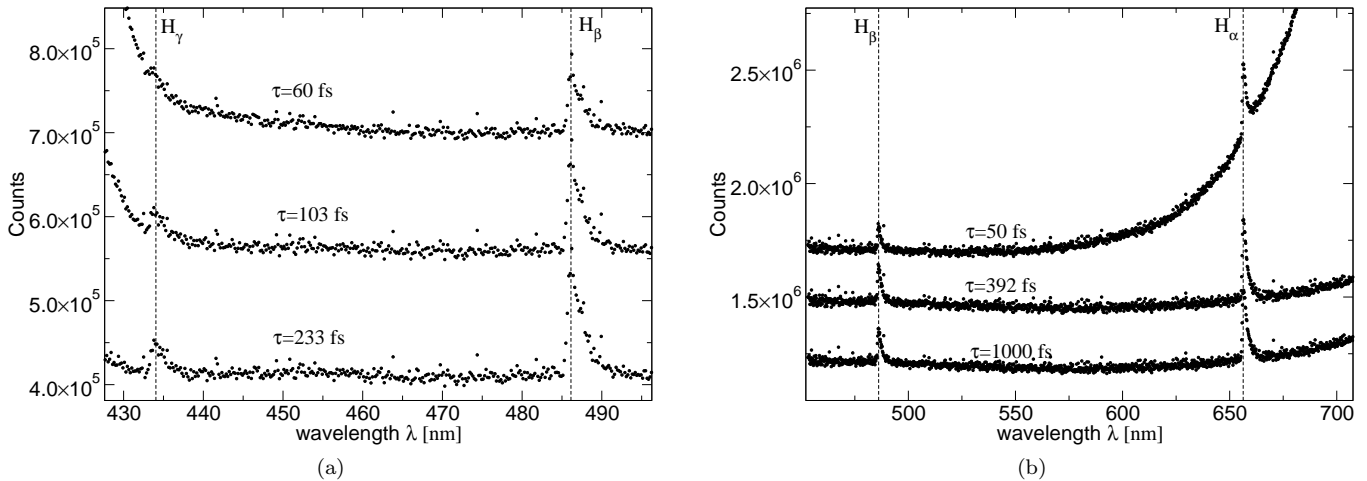

	    \subfigure[]{\includegraphics[width=.48\textwidth,angle=0,clip]{eps/H_spectrum_430-500_bw.eps}}
        \hfill
	\subfigure[]{\includegraphics[width=.48\textwidth,angle=0,clip]{eps/H_spectrum_460-700_bw.eps}}
    \caption{Representative emission spectra from laser-excited
    hydrogen microplasma in two spectral windows for various pulse
    widths $\tau$. The spectra are shifted by $1.5\cdot 10^{5}$ counts (a)
    and $2.5\cdot 10^{5}$ counts (b) with
    respect to each other.
     H Balmer spectral lines H$_\alpha, 656.3\,\mathrm{nm}$,
     H$_\beta, 486.1
    \,\mathrm{nm}$ and H$_\gamma, 434.1\,\mathrm{nm}$
    are given.
    The large enhancement at long wavelengths
    ($\lambda>660\,\mathrm{nm}$) is due to
    Rayleigh scattering of the laser light. 90000 shots are integrated
    at 1 kHz repetition rate.}
    \label{fig:H_spectra}
\end{figure}

The light is analyzed by a 0.34\,m
grating spectrometer (200 lines/mm) and monitored with a CCD
camera, providing a spectral dispersion of 0.185\,nm per pixel.
The entrance slit width was optimized for a high signal when
integrating 90000 shots at 1 kHz repetition rate. The resulting
spectral resolution of the spectrometer 
is $\Delta\lambda\simeq 1\,\mathrm{nm}$. Two regions
of the visible spectral range are analyzed in more detail, i.e.
(a) $430\ldots500$\,nm, and (b) $450\ldots 700$\,nm.
Fig.~\ref{fig:H_spectra} shows representative emission spectra from the
laser excited hydrogen microplasmas. Elastic Rayleigh scattering of the
incident radiation produces large enhancement of the measured
spectra towards the excitation wavelength $\lambda_0$. On top of
the background signal, H-Balmer spectral lines are clearly
identified, i.e.  H$_\alpha$ at $\lambda=656.3\,\mathrm{nm}$,
H$_\beta$ at $\lambda= 486.1 \,\mathrm{nm}$, and H$_\gamma$ at
$\lambda= 434.1\,\mathrm{nm}$.
\subsection{Analysis of the spectra\label{subsec:exp_hydrogen}}
\subsubsection{Effective Temperatures\label{subsec:effective_temp}}
\begin{figure}[ht]
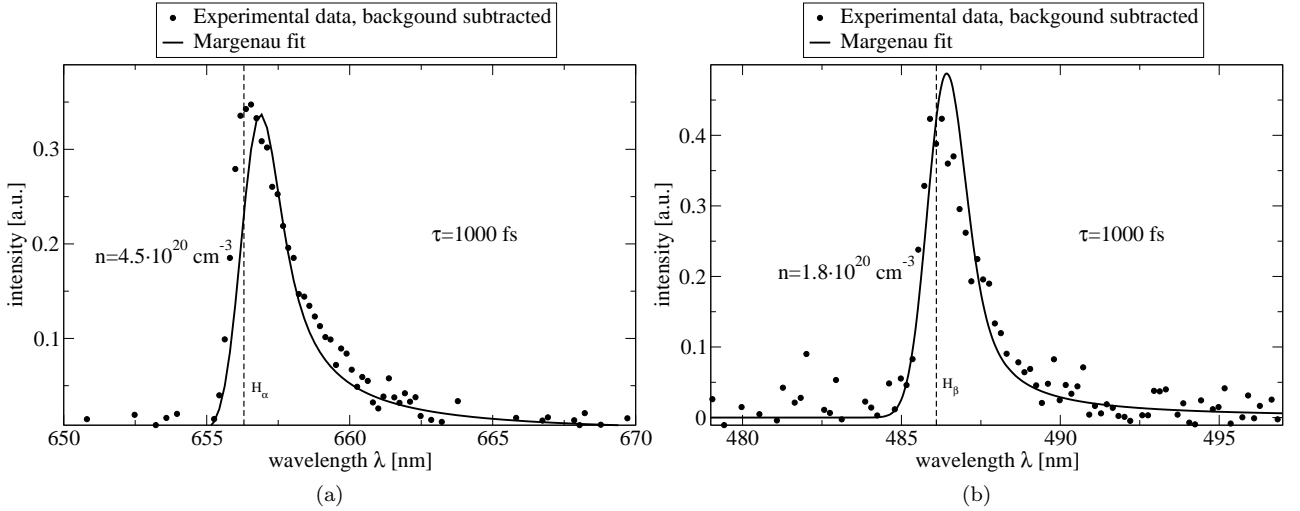

	\begin{center}
		\subfigure[]{\includegraphics[width=.48\textwidth,angle=0,clip]{eps/margenaufit_Ha_45e19.eps}}%
		\subfigure[]{\includegraphics[width=.48\textwidth,angle=0,clip]{eps/margenaufit_Hb_18e19.eps}}
	\end{center}
	\caption{Comparison of experimental line profiles (dots) of H$_\alpha$ and H$_\beta$ from the
	$\tau=1000\,\mathrm{fs}$ data to Margenau theory
	Eq.~(\ref{eqn:profile_london}), taking into account the
	limited detector resolution $\Delta\lambda\simeq
	1\,\mathrm{nm}$. For H$_\alpha$, good
	agreement between theory and experiment was obtained 
	setting $n^{(\alpha)}_\mathrm{atom}=4.5\cdot 10^{20}\,\mathrm{cm}^{-3}$. 
	For H$_\beta$, the best fit was obtained with
	$n^{(\beta)}_\mathrm{atom}=1.8\cdot
	10^{20}\,\mathrm{cm}^{-3}$.}
	\label{fig:1000fs_H_a_Margenaufit}
\end{figure}
In order to quantitatively analyze the Balmer line signals, 
one has to subtract the
background signal. 
Therefore,  the neighbouring
spectral ranges of each line, i.e. 640..655 nm
and 670..685 nm for the H$_\alpha$ line,
 460..467 nm and 472..480 nm for H$_\beta$ as well as 430..433 nm and 
 438..445 nm for H$_\gamma$ are fitted by an 
 exponential function, which is then subtracted from the data
 set.
As an example,  Figs.~\ref{fig:1000fs_H_a_Margenaufit} (a) and (b) show 
the background subtracted spectra for a pulse width of $\tau=1000$ fs
(dots). The solid curves are the Margenau profiles fitted to the data;
they will be discussed in more detail in
Sec.~\ref{subsec:vdw-broadening}. 

By comparing integrated intensities from different spectral lines, one
can determine the effective temperature in the system.  Given the
ratio of wavelength-integrated intensities $I_1/I_2$ of two spectral
lines of the same atomic species and the same ionization stage at
central wavelengths $\lambda_1$ and $\lambda_2$, and assuming again
local thermal equilibrium, the effective temperature is
\cite{lochte-holtgreven_in_LH}
\begin{equation}
	T^\mathrm{eff}_{12}=\frac{E_1-E_2}{k_\mathrm{B}\,\ln\left[ \frac{A_1\,
              g_1\,\lambda_2}{A_2\,g_2\,\lambda_1}\frac{I_2}{I_1}
          \right]}~. 
	\label{eqn:Teff}
\end{equation}
$A_1$, $A_2$ are the angular momentum averaged Einstein coefficients
for the transition between levels corresponding to the spectral line 1
and 2, respectively and $g_1$ and $g_2$ are the degeneracy factors of
the excited level (principal quantum number
$n_i^\mathrm{ex}$) of each transition, i.e. $g_i=2(n^\mathrm{ex}_i)^2$.
$E_1$ and $E_2$ are the energies of the excited level for each
transition.  Tab.~\ref{tab:Teff} lists effective temperatures for each
pulse width.  As can
be seen, $T^\mathrm{eff}_{12}$ is
dependent on $\tau$. One obtains different results from
the two ratios $I(\textrm{H}_\alpha)/I(\mathrm{H}_\beta)$ and
$I(\textrm{H}_\beta)/I(\mathrm{H}_\gamma)$.  For small $\tau$,
$T^\mathrm{eff}_\mathrm{\alpha\beta}$ rises with increasing $\tau$. At $\tau\simeq
400\,\mathrm{fs}$, $T^\mathrm{eff}_{\alpha\beta}$ saturates near $1\,\mathrm{eV}$.
On the other hand, $T^\mathrm{eff}_{\beta\gamma}$ gives values around
$0.5\,\mathrm{eV}$. 

\begin{table}[!h]
	\centering
	\caption{Effective line temperatures according to
	Eq.~(\ref{eqn:Teff}) versus pulse length. The first set of
	figures was obtained from the short wavelength spectrum,
	shown in Fig.~\ref{fig:H_spectra}(a), the second set 
	results from the data shown in Fig.~\ref{fig:H_spectra}(b).}
	\begin{tabular}{|c|ccc||c|ccccccc|}
		\hline
		$\tau [fs]$&60&103&233&
		$\tau [fs]$&50&109&170&264&392&695&1000\\
		\hline
		$T^\mathrm{eff}_\mathrm{\beta\gamma}$ [eV]& 0.49&
		0.56& 0.41&
		$T^\mathrm{eff}_\mathrm{\alpha\beta}$ [eV]& 
		0.57& 0.62& 0.81& 0.97& 0.99& 1.00& 1.04\\
		\hline
	\end{tabular}
	\label{tab:Teff}
\end{table}
These effective temperatures have to be analyzed within a microscopic
description of the expanding plasma source. Self-absorption should be
considered so that the intensities occurring in Eq.~(\ref{eqn:Teff})
are modified when propagating through the microplasma.
Furthermore, different values $T^\mathrm{eff}_{12}$ from the intensities of
H$_\alpha$, H$_\beta$ and H$_\gamma$ indicates that the system is in a
generic non-equilibrium state, which cannot be described by a single
temperature. Despite the measured spectra are time integrated, we have
to assume a dynamic evolution of the plasma parameters temperature and
density over the course of the expansion of the excited hydrogen
droplets.  Within a hydrodynamical description given below in
Sec.~\ref{subsec:hydro}, local thermal
equilibrium with plasma parameters depending on space and time are
introduced.

During the expansion process, radiation can only be produced if the
upper level of the transition in question is existent in the system so
that it can be occupied. From simple geometric and energetic arguments
it is clear, that $n=3,4,5$ levels are not present in hydrogen at
liquid densities. This point will be discussed in more detail in
Sec.~\ref{subsec:levels}.

The continuum background stems from free-bound and free-free
(bremsstrahlung) transitions. The spectral behavior can also be used
to estimate effective temperatures which should be analyzed within a
dynamical expansion model. Since more experimental work is necessary
to separate the continuum background radiation, we will consider only
the line spectra in this work.
\subsubsection{Spectral line shapes\label{subsec:vdw-broadening}}
To analyze the line profiles, one first has to consider the
effect of the broadening due to the finite resolution  
of the experimental setup. The true line profile is obtained after 
deconvolution with the detector function.
In a separate
experiment the spectral resolution was measured
as $\Delta\lambda\simeq 1\,\mathrm{nm}$ .
This is consistent with the steep increase of intensity on the
blue wing of the unperturbed spectral line.

\begin{figure}[ht]
	\begin{center}
		\includegraphics[width=.8\textwidth,angle=0,clip]{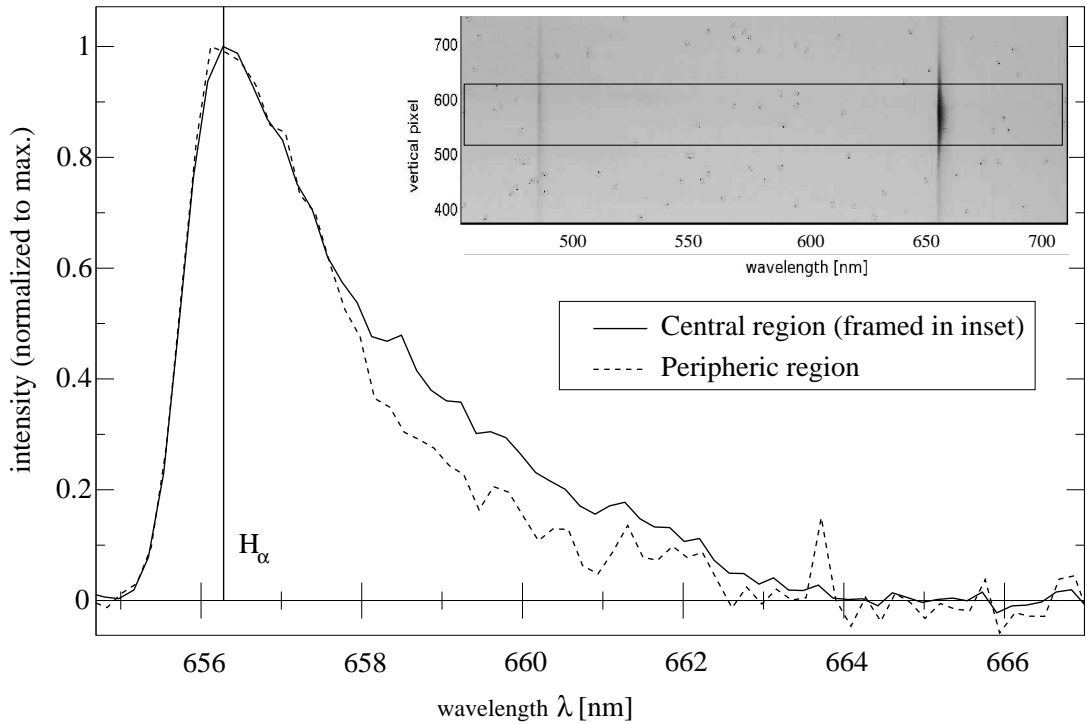}
	\end{center}
	\caption{Normalized spectra near the $\mathrm{H}_{\alpha}$-line for
	$\tau=1000$ fs pulse length. The raw data (cf. inset) have
	been summed over different vertical ranges. The solid line
	marks the central region of the laser focus (framed part in the inset), the dashed curve corresponds to the data from the
	peripheric region.
	}
	\label{fig:h_1000fs_2cuts}
\end{figure}

On the long wavelength wing, all spectral lines show a significant
broadenening. 
In Fig.~\ref{fig:h_1000fs_2cuts} 
the spectral line shape is analyzed as a function of the distance of the
origin of radiation from the plasma's center. This is achieved by
integration of the CCD-Data over different ranges in the
vertical direction, cf. the inset in Fig.~\ref{fig:h_1000fs_2cuts}.
The corresponding
spectra for the central region 
of the laser focus (framed part in the inset) and for the
peripheric region, about 100 $\mu$m beside the focus are shown. 
After background
subtraction, the spectra are
normalized to their maximum value. In this way it can be seen, that
the width of the spectral line depends significantly on the spatial
origin of the radiation and decreases with increasing distance from the
center. This behaviour indicates that the asymmetric broadening of the
spectral lines is related to the properties of the 
microplasma and should be described as a density effect.

Pressure broadening of spectral lines is caused by charged 
as well as by neutral perturbers. As discussed in
Sec.~\ref{sec:theory_lineshapes}, free
electrons are treated in impact approximation leading to a symmetric
Lorentzian line profile. Also the ionic microfield contributes to a
Voigt profile with symmetric  broadening on both the red as well as
the blue wing (linear Stark effect). Both effects are smaller than
the spectral resolution of 1\,nm, as can be seen in the spectra, e.g.
Fig.~\ref{fig:h_1000fs_2cuts}. Thus,
we conclude that the influence of free charged
particles is not clearly identified so that the free electron
density is below $n_e^{\rm free}=10^{17}\,\mathrm{cm}^{-3}$ 
which at $T$ = 1 eV gives for H$_\alpha$ the
FWHM of 1 nm.

The asymmetric red shift can be described by the interaction with
neutral perturbers.  We will now compare the experimental data to the
line-shape due to interaction with neutral perturbers, as outlined in
Sec.~\ref{sec:bound-bound_born}.  In
Fig.~\ref{fig:1000fs_H_a_Margenaufit}, the measured spectra
in the vicinity of the H$_\alpha$ line are compared to the emission profile as
given by Margenau, Eq.~(\ref{eqn:profile_london}), convoluted with the
detectors resolution function \cite{gaussian}. In the case of the H$_\alpha$-line, the
best fit was obtained using $n^{(\alpha)}_\mathrm{atom}=4.5\cdot
10^{20}\,\mathrm{cm}^{-3}$ for the density of neutrals. In the case of
H$_\beta$, the density $n^{(\beta)}_\mathrm{atom}=1.8\cdot 10^{20}\,\mathrm{cm}^{-3}$
gives the minimum $\chi^2$.

As for the effective temperatures, also the different values for the
effective density
contradict an equilibrium
picture. A dynamical description of the microplasma's evolution
is needed. This task will be accomplished in the next section by means
of hydrodynamical simulations.
\section{Dynamical plasma expansion model\label{sec:discussion}}
In the previous section the effective temperature and density of the
microplasma were determined by analysis of the experimental spectra.
For both quantities, different values  have been obtained, depending on
the external parameters, e.g. the pulse width of the laser, and on
the wavelengths considered in the analysis. This strongly indicates
that the plasma parameters are time-dependant quantities, which have
to be modeled, e.g. by hydrodynamic simulations to improve the
description of the microplasma and the interpretation of the
experimental data.

\subsection{Hydrodynamic expansion\label{subsec:hydro}}

Hydrodynamic simulations are a versatile tool to infer
the dynamics of a strongly coupled many-particle system.
In the case of plasmas, hydrocodes have been
successfully applied to study these systems under the influence of 
strong external fields \cite{Eidmann_PhysRevE.62.1202} as well as the 
relaxation of a plasma in an
excited state into equilibrium.
Here, we will study the
hydrodynamic expansion of the excited, i.e. heated H microdroplet 
after times which are
long compared to the pulse length of the laser. 
Simulations are performed using
the hydrocode MULTI2002 
\cite{Ramis_CompPhysComm49_474_1988,Ramis_NuclFus44_720_2004}. 
As initial conditions, we assume a homogenous
density profile $n(r,t=0)=n_\mathrm{liq}=4.5\cdot
10^{22}\,\mathrm{cm}^{-3}$ and also homogeneous temperature
distribution.  Assuming
50 \% absorption of the laser energy given above by the droplet
\cite{Zweiback_PRL84_2634_2000} and accounting for the
ratio of the droplet's cross-section to the focal spotsize, i.e.
(10\ $\mu$m/40\ $\mu$m)$^2\simeq 0.06$ as well as for the energy needed to
break up the molecular bounds, i.e. 2.2 eV per atom, we obtain
$T(r,t=0)=14\,\mathrm{eV}$ as the initial temperature.
\begin{figure}[ht]
	\begin{center}
		\includegraphics[width=.48\textwidth,angle=0,clip]{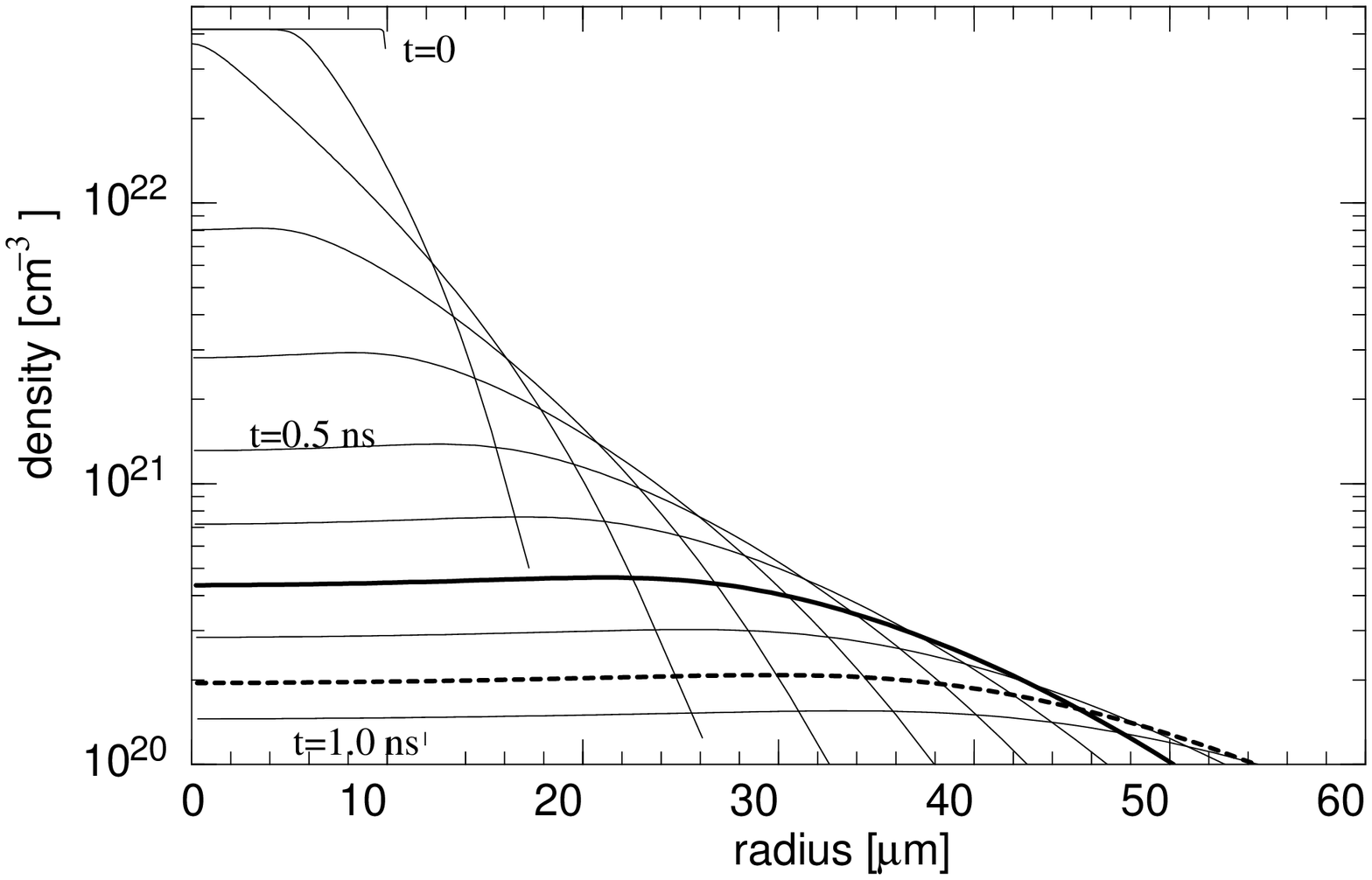}
		\hfill
		\includegraphics[width=.48\textwidth,angle=0,clip]{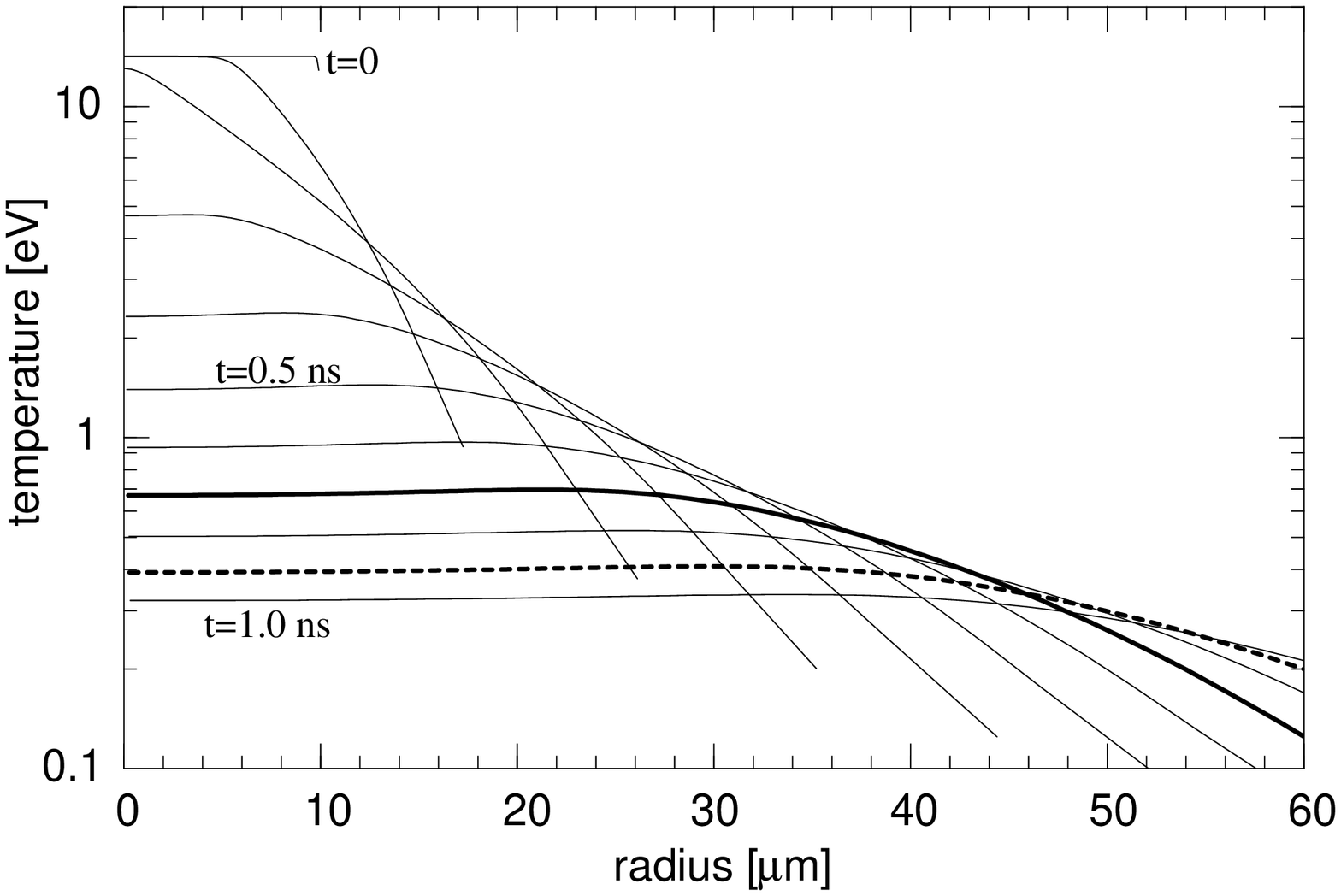}
	\end{center}
	\caption{Hydrodynamic simulation of the density (left) and
	temperature (right) profiles of an 
	expanding hydrogen microplasma
	starting from $n_\mathrm{liq}=4.5\cdot 10^{22}\,\mathrm{cm}^{-3}$ and
	$T=14\,\mathrm{eV}$. The broad solid and dashed curves give
	the conditions after 0.7 ns and 0.9 ns. At this time, the
	central density has reached the density which has been obtained
	from the fit of the Margenau profile to the H$_\alpha$ and 
	the H$_\beta$ spectral line, respectively for $\tau=1000\,\mathrm{fs}$
	pulselength. The corresponding
	temperatures at these times are close to the effective temperatures
	$T^\mathrm{eff}_{\alpha\beta}\simeq 1\,\mathrm{eV}$ and
	$T^\mathrm{eff}_{\beta\gamma}\simeq 0.5\,\mathrm{eV}$ obtained
	from the spectra, see Tab.~\ref{tab:Teff}.
	}
	\label{fig:dens-temp_profile}
\end{figure}

\begin{figure}[ht]
	\begin{center}
		\subfigure{\includegraphics[width=.48\textwidth,angle=0,clip]{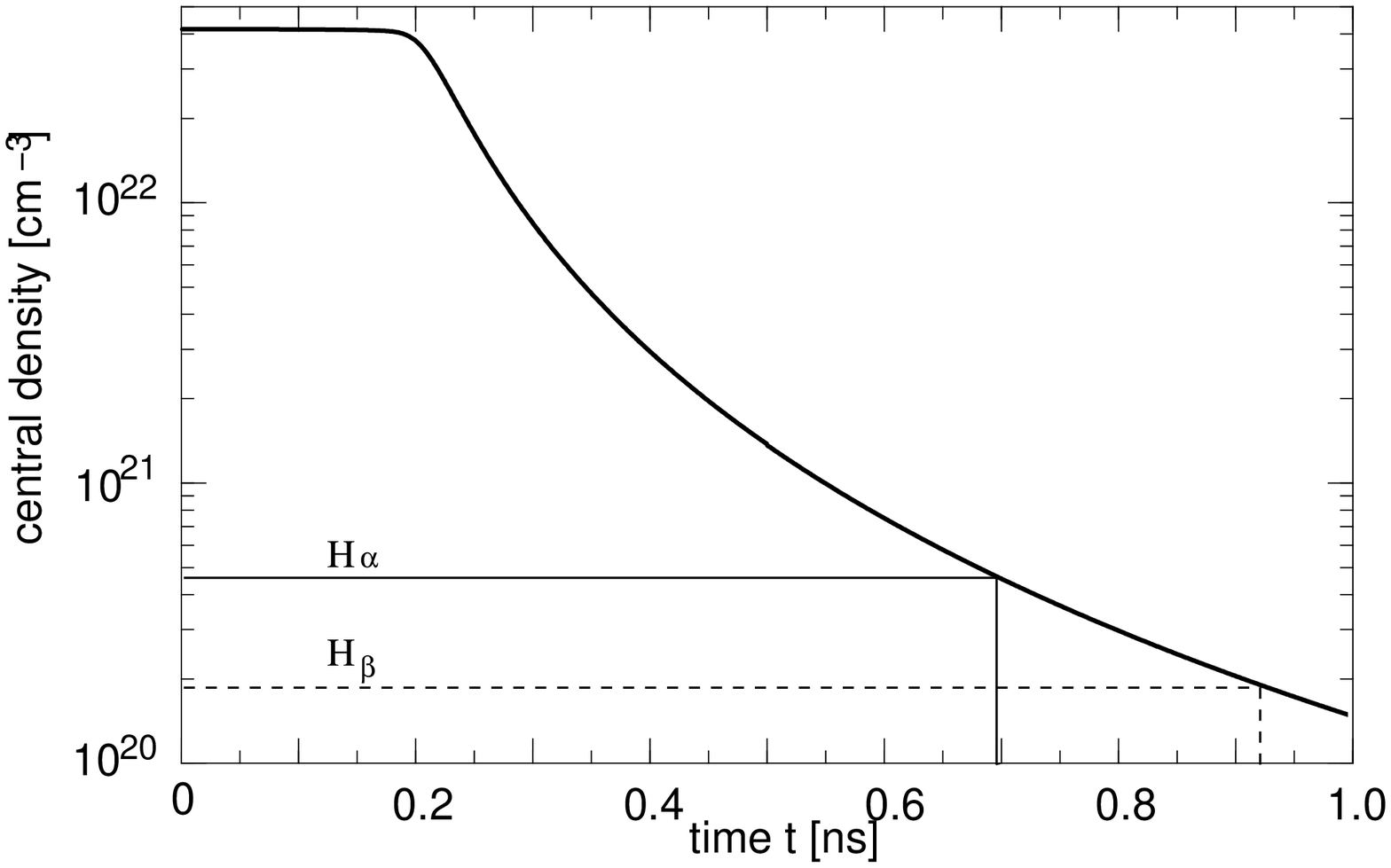}}
		\hfill
		\subfigure{\includegraphics[width=.48\textwidth,angle=0,clip]{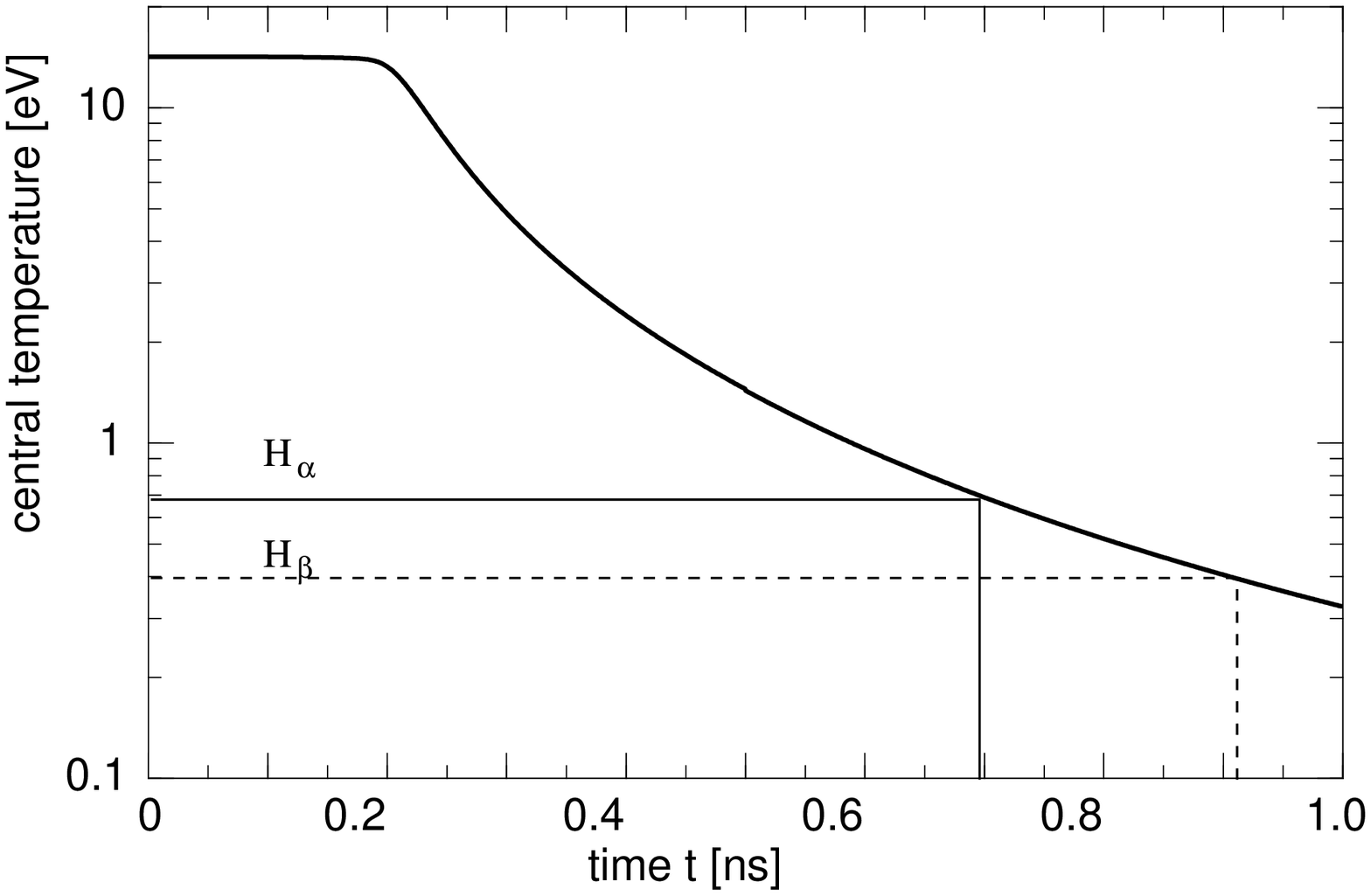}}
	\end{center}
	\caption{Temporal evolution of central density $n(r = 0, t)$
	(left)
	and temperature $T(r = 0, t)$ (right), given as bold solid lines. Narrow solid line:
	central parameters after 0.7 ns, i.e. at the time, when the density
	obtained in the fit of the Margenau profile to the H$_\alpha$
	line, $n^{(\alpha)}_\mathrm{atom}=4.5\cdot 10^{20}\,\mathrm{cm}^{-3}$ is reached.
	Dashed line: the same for H$_\beta$.}
	\label{fig:dens-temp_evolution}
\end{figure}

Fig.~\ref{fig:dens-temp_profile} shows the
number density and temperature profiles. The uppermost
curve corresponds to $t=0\,\mathrm{ns}$, while the last curve is
taken at $t=1\,\mathrm{ns}$. Both parameters $n(r,t)$ and $T(r,t)$
show a fast decline in the center
of the droplet. The density decreases by a factor of 10 every 300
ps. Secondly, 
all profiles are nearly constant over several tens of micrometers
starting from
the center and decrease sharply when approaching the rim of the
droplet. Therefore, the central values $n(r=0,t)$ and $T(r=0,t)$ 
can be taken as representative values for the whole droplet. 
The temporal evolution of
the central number density and temperature is shown in Fig.
\ref{fig:dens-temp_evolution}. 
\begin{figure}[ht]
	\begin{center}
		\includegraphics[width=.5\textwidth,angle=0,clip]{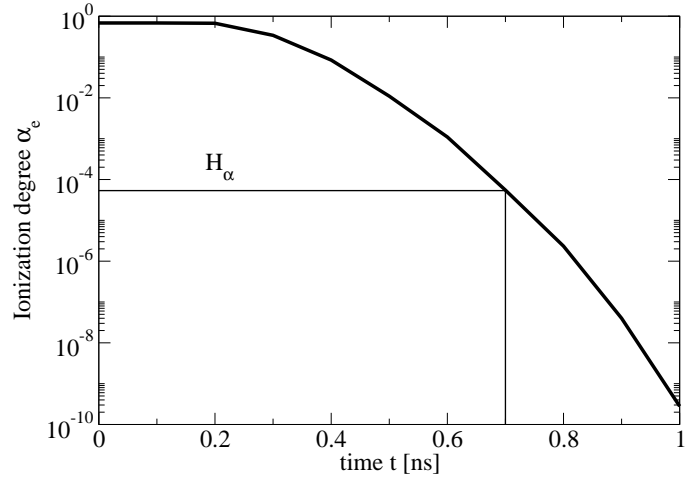}
	\end{center}
	\caption{Ionization degree $\alpha_e$ as a function of time
	obtained from Saha equation (bold solid curve). The narrow 
	horizontal line gives the
	ionization after 0.7 ns, i.e. at the time, when the central
	density has decreased to $n^{(\alpha)}_\mathrm{atom}=4.5\cdot
	10^{20}\,\mathrm{cm}^{-3}$, which has been obtained as
	best-fit parameter with respect to the Margenau profile of the 
	H$_\alpha$ spectral line.}
	\label{fig:saha}
\end{figure}

\newpage

The hydrodynamical description gives the time evolution of
the excited droplet. The assumption of local thermal equilibrium may
be justified on the time scale of ps for the thermalization of kinetic
energies. If assuming also local ionization equilibrium, the composition
during the time evolution can be calculated. 
The temporal behaviour of the ionization in the droplet's center
is plotted in 
Fig.~\ref{fig:saha}.
For the initial conditions $T=14\,\mathrm{eV}$ and
$n_\mathrm{liq}=4.2\cdot
10^{22}\,\mathrm{cm}^{-3}$, we have an ionization degree
$\alpha_e=n_\mathrm{free}/n=0.68$. 
After $t=0.7\,\mathrm{ns}$, the 
density and temperature have decreased so far, that $\alpha_e$ 
drops below $10^{-4}$. This corresponds to a concentration of
free electrons of roughly $10^{17}\,\mathrm{cm}^{-3}$. At this
density, and $T\simeq 1\,\mathrm{eV}$, the Stark effect leads to a 
broadening of the H$_\alpha$ line of 1 nm, as was discussed in
Sec.~\ref{subsec:vdw-broadening}. Thus, at times larger than 0.7 ns,
Stark broadening does not give a notable contribution to the width of the
spectral lines. 

This analysis has shown, that  the emission of Balmer line
radiation occurs at comparatively late times of the expansion, when
the density has decreased to a certain level. This is due to the fact,
that in a dense system, excited levels are not well defined due to
interaction with neighbouring atoms. Both wavefunctions and
atomic potentials are disturbed. In the following section we will 
analyse the question, which density has to be established in the
system, so that excited atomic levels are defined and the
corresponding radiative transition may occur.

\subsection{Occupation of excited levels\label{subsec:levels}}
In a dense system, the potential energy of an electron in a given atom
is modified by the medium \cite{Siedschlag_PRA71_031401_2005}. Thus, energy eigenvalues and wave functions
are changed due to screening by free charge carriers. In particular,
it is well-known that bound states merge with the continuum of
scattering states at high densities so that the electrons are no
longer bound to a special ion, but move relatively free in the plasma
as denoted by a transition from dielectric to metallic
behavior \cite{krae}. 
Considering excited states, the dissolution occurs already at
lower densities so that these excited states cannot any longer be occupied
to produce a line spectrum.

\begin{figure}[ht]
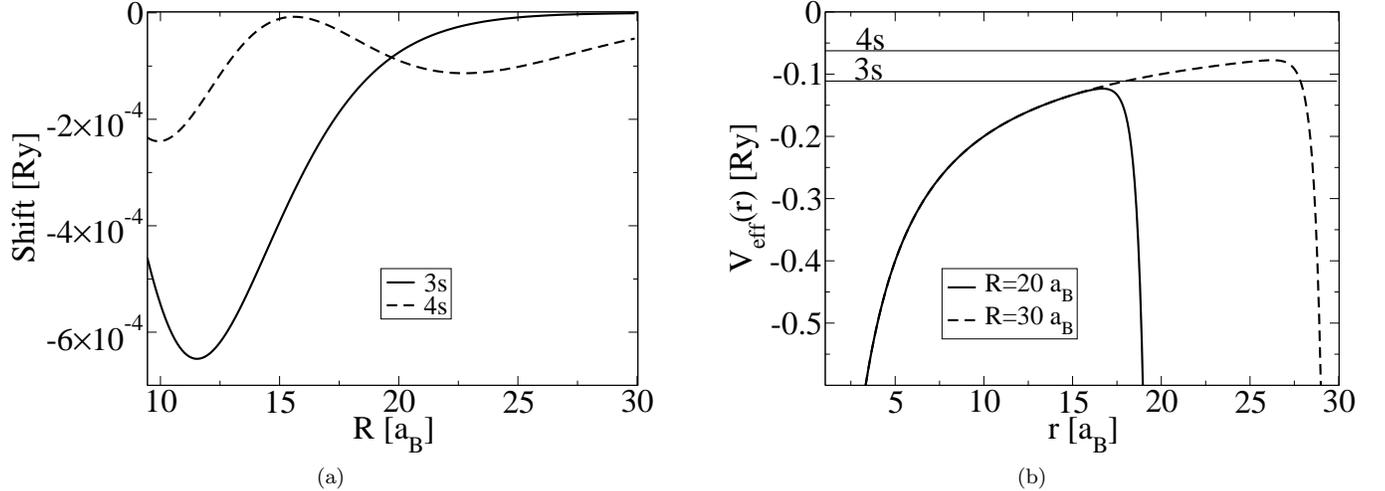

 	\begin{center}
	\subfigure[]{%
	\includegraphics[width=.48\textwidth,angle=0,clip]{eps/potential-shifts_3s-4s_bw.eps}}
	\hfill
	\subfigure[]{%
	\includegraphics[width=.48\textwidth,angle=0,clip]{eps/shifts_R_3s-4s.eps}}
 	\end{center}
 	\caption{(a) Energy shift of 3s and 4s levels as a function of
	the distance $R$ to the next neighbouring
	atom. \\
	(b) Effective potential due to the overlap between the radiator's Coulomb potential and the screened potential of the next
 	neighbouring atom in its ground state. Results are shown for two different values of the interatomic distance $R=20a_\mathrm{B}$
 	(bold curve) and $R=30\,a_\mathrm{B}$ (dashed curve).  The
	perturbed 3s ($R=20\,a_\mathrm{B}$) and
	4s ($R=30\,a_\mathrm{B}$) energies are given as narrow
	lines.\\
}
 	\label{fig:potential_shifts}
 \end{figure}

We focus here to the influence of bound states in the medium to
analyze at which densities the excited states are dissolved into the
continuum so that no line spectra are formed. In a first
approximation, we calculate the modification of the potential in the
Schr\"odinger equation for the hydrogen atom due to the perturbing
atoms in the surrounding medium. Only ground state atoms are
considered, 
and the correlations between the electrons in the perturbing
ground state with the radiating electron in the excited state are
neglected. Solving
the Poisson equation, the effective potential is obtained as
superposition of the Coulomb potential and the potential of the
neighboured atoms, which leads to a lowering of the Coulomb
potential by the amount
\begin{equation}
\Delta V_i(\vec r) = -{e^2 \over 4 \pi \epsilon_0 |\vec r - \vec R_i|}
e^{-2 |\vec r - \vec R_i|/a_B} \left(1 + {|\vec r - \vec R_i| \over
    a_B} \right)\,.
\end{equation}
Here, $\vec R_i$ is the position of the perturbing atom.
The lowering of the Coulomb potential has two effects: On the one
hand, the energy level is shifted to lower energies, on the other
hand, at a certain critical distance, the threshold of potential
energy between neighboured atoms crosses the
shifted electrons energy level, and the bound state is dissolved into
the continuum. This is the case, if the distance between
neighbours comes into the range of the extension of the electrons
wave-function.  
We assume a closely packed configuration where the radiating atom is
surrounded by 12 perturbing next neighbours. The shift of the energy
levels with principal quantum numbers $n$ = 3 and 4 has been evaluated
in first order of perturbation theory and can be given in an
analytical form. Results for the shift as a function of the 
interatomic distance are given in Fig.~\ref{fig:potential_shifts}(a).

In Fig~\ref{fig:potential_shifts}(b) we give the potential energy $V(r)$ in the direction to a
next neighbour for two different distances, together with the
corresponding shift of the energy levels with $n$ = 3 and $n=4$. As can be seen,
the bound state disappears at a critical distance $R_c^{(n)}$ where
the threshold energy becomes lower than the binding energy, and the
electron is no longer bound to the central ion, but escapes the
effective potential and moves relatively freely within the cluster.
 
From the critical distance $R_c^{(n)}$, one can then infer the
``critical'' density $n_c^{(n)}$ that has to be
established in the expanding
system, before the upper level $n$ of the considered transition can exist,
i.e. before the corresponding spectral line can occur.  In this way,
we obtained for the second excited level, relevant for the H$_\alpha$
transition, a ``critical'' density of $n^{(\alpha)}_c\simeq 1.24\cdot
10^{21}\,\mathrm{cm}^{-3}$.  This
value for $n_c$ is by a factor of two larger than the value
$n^{(\alpha)}_\mathrm{atom}=4.5\cdot 10^{20}\,\mathrm{cm}^{-3}$ obtained from the fit of the
$H_\alpha$ line to the line shape due to van der Waals interaction,
cf.  Fig.~\ref{fig:1000fs_H_a_Margenaufit}(a). 

It has to emphasized in this context, that these are exploratory
calculations to estimate the region of density where well-defined
excited energy levels of the radiating atom can exist. A more detailed
calculation should include electronic correlations between radiator
and perturber which have been neglected in the calculation of the
perturbing potential. The account for correlation and exchange effects
will modify the potential and the critical density. In addition, we
considered only a perturbing atom at mean distance, neglecting any
fluctuations in configurations. On the other hand, the experimental 
spectra are time integrated
measurements and hence the density has to be interpreted as a mean
value, averaged over the whole exposure time. At the moment the
radiation starts, the density might in fact be larger
than the value inferred from the measured spectra.  For $H_\beta$, the
critical density is obtained as $n_c^{(\beta)}\simeq 2.66\cdot
10^{20}\,\mathrm{cm}^{-3}$, which also exceeds the fit parameter for
the H$_\beta$ line ($n^{(\beta)}_\mathrm{atom}=1.8\cdot 10^{20}\,\mathrm{cm}^{-3}$) 
for the same reasons as given above for the case of the H$_\alpha$
line (cf.  Fig.~\ref{fig:1000fs_H_a_Margenaufit}(b)).

In conclusion, the experimental Balmer spectra are not characteristic
for
the
first stage of the laser excitation of the cluster, because the
density of hydrogen in the condensed state is too high to form
well-defined excited atomic levels with the corresponding principal
quantum numbers. Such levels appear only during the process of
expansion and may be used as a signal to infer the state of the
microplasma at the corresponding time stage.

\subsection{Line emission scenario\label{subsec:scenario}}

As discussed in the previous Section, the observed Balmer spectra
cannot be interpreted within an equilibrium picture of the
laser produced microplasma. We could not infer consistent values of
plasma parameters for temperature and density. A consistent description
is only possible if the time evolution of the expanding microplasma is
considered as a non-equilibrium process.

We follow the dynamical expansion of the laser excited hydrogen
droplet as given by the hydrodynamical calculation. First, 
the time evolution of the density is considered, with account to
the
effective densities derived from the measured line profiles. 
Thereby, one can 
determine the time at
which the H$_\alpha$ line appears, i.e. where the inferred density  
$n^{(\alpha)}_\mathrm{atom}=4.5\cdot 10^{20}\,\mathrm{cm}^{-3}$ is reached. Starting with the
excitation temperature $T = 14$ eV after the short-pulse laser
excitation, this density is established after 0.7 ns. The corresponding
temperature obtained from the hydrocode MULTI is $T=0.65$ eV and is in
reasonable agreement with the estimate for
$T^\mathrm{eff}_{\alpha\beta}$ given in
Sec.~\ref{subsec:vdw-broadening}.
Thus, the  H$_\alpha$ line profile reflects the state of the
expanding microplasma after 0.7 ns.
In the case of the H$_\beta$ we observe the following:
From the van der Waals fit to the data, cf.
Fig.~\ref{fig:1000fs_H_a_Margenaufit}(b), $n^{(\beta)}_\mathrm{atom}=1.8\cdot
10^{20}\,\mathrm{cm}^{-3}$ was obtained, which is reached in the
hydrodynamic simulation after $t\simeq 0.9\,\mathrm{ns}$. The
temperature is $T\simeq 0.4\,\mathrm{eV}$ at that moment, which
is close to the value $0.5\,\mathrm{eV}$ 
obtained for $T^\mathrm{eff}_{\beta\gamma}$, cf. Tab.~\ref{tab:Teff}.
H$_\beta$ thus gives us information about the state of the droplet at
$t\simeq 0.9\,\mathrm{ns}$.

Second, as for the empirically determined plasma parameters, the critical
densities for H$_\alpha$ and H$_\beta$ as discussed in
Sec.~\ref{subsec:levels},
have to be compared to the results of the hydrodynamic
simulation. 
The central density of the droplet decreases to the critical density
for H$_\alpha$ to appear, $n_c^{(\alpha)}=1.24\cdot
10^{21}\,\mathrm{cm}^{-3}$ after 0.45 ns. After the same time, the
temperature at the center of the droplet has decreased to 1.3 eV,
which is in good agreement with the value 1 eV inferred from
$T^\mathrm{eff}_{\alpha\beta}$.
The critical density for H$_\beta$, $n_c^{(\beta)}=2.66\cdot
10^{20}\,\mathrm{cm}^{-3}$ is reached after 0.8 ns. The central 
temperature is 0.5 eV at that moment, which again coincides well
with the value 0.5 eV as obtained from $T^\mathrm{eff}_{\beta\gamma}$,
cf. Tab.~\ref{tab:Teff}. 

This analysis shows, that the droplet in fact undergoes a complex dynamical
evolution. Using hydrocodes to model the droplets history, we can
understand the experimental observations. 
The initial temperature of $14$ eV is consistent with the observed 
temperatures and
densities established at later times of the evolution. 
The line radiation obviously stems from relatively late times.
 What happens before line radiation appears?
At the initial conditions of $n_\mathrm{atom}=n_\mathrm{liq}=4.2\cdot
10^{22}\,\mathrm{cm}^{-3}$ and $T=14\,\mathrm{eV}$, the ionization of
the droplet was calculated as $\alpha_e=68\%$, solving Saha's equation, 
cf. Fig.~\ref{fig:saha}.
The plasma
emission is thus dominated by continuum radiation (free-free and free-bound
transitions). However, since the detector integrates over the whole
evolution of the plasma, only a nearly constant background remains in the
spectrum from this early stage of the evolution. As the ionization
decreases, bound states begin to form and line radiation occurs. However,
for a given spectral line, both upper and lower level of the
corresponding transition have to be well defined. To this end, the
density has to fall below a certain value, in order to allow for the
excited energy level to appear underneath the effective potential in the
dense medium.

Considering the time scales for the expansion of the cluster of the
order $10^{-9}\,\mathrm{s}$ according to the hydrodynamical calculations, the local
thermodynamic equilibrium can be assumed, and the life time of the
excited states of the order $10^{-11}\,\mathrm{s}$ is short compared with the
evolution of the microplasma. In a more rigorous approach, the
variation of temperature and density in space and time should be
accounted for to synthesize the spectra.

\section{Conclusions\label{sec:concl}}

We found that the visible spectra can be used to get signatures for a
time-resolved picture of an expanding microplasma. The line profiles 
allow for the determination of the microplasma's temperature and density
during its hydrodynamic expansion. 
We considered an energy deposition in the liquid hydrogen droplet
which is sufficiently weak so that in the expanded phase most of the
electrons are found in atomic bound states. Thus, the spectral line
profiles are determined by van der Waals broadening. We presented a
general quantum statistical approach to line profiles which allows for
the unified description of charged and neutral perturbers, including
density effects such as dynamical
screening or strong collisions.   

Although the spectra are time integrated measurements, 
we can use the Balmer lines
as signatures for the expanding microplasma at definite values of
density. A consistent scenario for the emission of Balmer lines has
been given. To follow the time evolution of an excited droplet more
directly, time resolved spectra have to be analyzed.  Possibly, this
can be achieved with pump-probe experiments and/or stimulated emission
measurements.
\section*{Acknowledgements}
This work has been supported by the Deutsche Forschungsgesellschaft
(DFG) through the 
Collaborative Research Center (SFB) 652. TL acknowledges financial
support from the DFG under Grant No. LA 1431/2-1.



\begin{thebibliography}{52}
\bibitem[{not()}]{note_groundstate}
\bibinfo{note}{To keep the notation short, we denote the ground state of an
  atom by $\alpha=0$. For hydrogen atoms, this has to be read as $nl=1s$.}
\bibitem[{gau()}]{gaussian}
\bibinfo{note}{We use the normal distribution
  $D(\lambda)=\exp(-\lambda^2/2\sigma^2)/\sqrt{2\pi}\sigma$ with
  $\sigma=0.425\cdot \Delta\lambda$, such that $\Delta\lambda$ is the full
  width at half maximum of the distribution $D(\lambda)$.}
\end{thebibliography}


\begin{appendix}

\section{Green function approach to the polarization
function\label{app:pi_2}}

The general expression for the single state contribution $\Pi_1(\vec
k, \omega)$ to
the polarization function has the form
\begin{equation}
	\Pi_1(\vec k,\omega_\mu)=
	\parbox{80pt}{%
	\psfig{figure=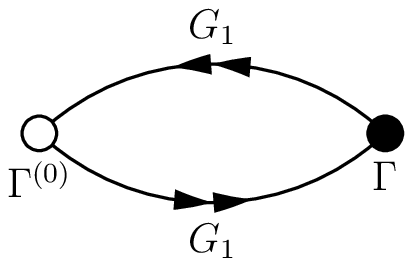,width=80pt}}~.
	\label{eqn:pi1_diagram}
\end{equation}
The full single-particle propagator contains the self-energy
$\Sigma_1(\vec p, z)$,
\begin{equation}
	\parbox{80pt}{\psfig{figure=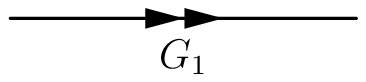,width=80pt}}\,=\,
	\parbox{80pt}{\psfig{figure=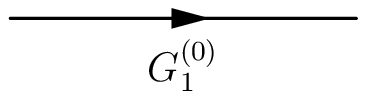,width=80pt}}\,+\,
	\parbox{80pt}{\psfig{figure=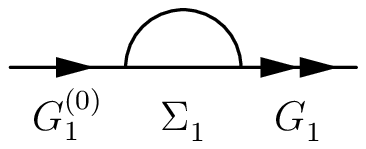,width=80pt}}~.
	\label{eqn:dyson_diagram}
\end{equation}
The vertex $\Gamma$ describes the coupling to the
electromagnetic field and can also be expressed in terms of an
effective interaction kernel. 
Both quantities, self-energy $\Sigma$ and vertex $\Gamma$,
have to be approximated
in a consistent way. For the self-energy one has the 
$GW\Gamma$ approximation, given by the diagram
\begin{equation}
	\Sigma^{GW\Gamma}(\vec p,z_\nu)=\parbox{80pt}{%
	\psfig{figure=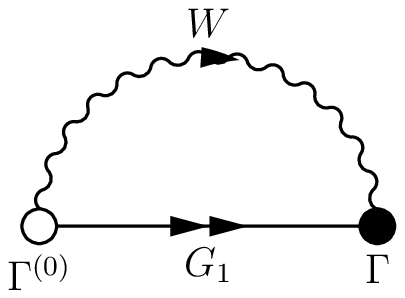,width=80pt}}~,
	\label{eqn:dse_full}
\end{equation}
which is a self-consistent equation for the full propagator, the full
vertex $\Gamma$, as well as for the
screened interaction potential $W$,
\begin{align}
	W(\vec k,\omega_\mu)&=
	\parbox{50pt}{%
	\psfig{figure=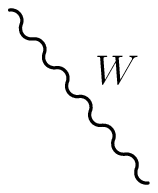 ,width=50pt}}=
	\parbox{40pt}{%
	\psfig{figure=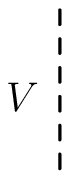 ,width=40pt}}+
	\parbox{80pt}{%
	\psfig{figure=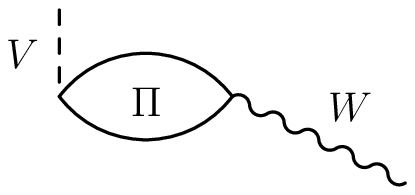 ,width=80pt}}
	=\frac{V(\vec k)}{1-V(\vec k)\Pi(\vec k,\omega_\mu)}~,
	\label{eqn:Vscreened_diagrams}
\end{align}
while the vertex function $\Gamma$ is the solution of the Bethe-Salpeter
equation, given in diagrammatic form by
\begin{equation}
	\parbox{40pt}{\psfig{figure=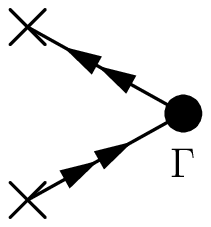,width=40pt}}\,\,\,=\,
	\parbox{40pt}{\psfig{figure=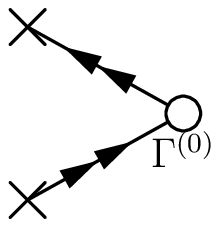,width=40pt}}\,\,\,+\,
	\parbox{80pt}{\psfig{figure=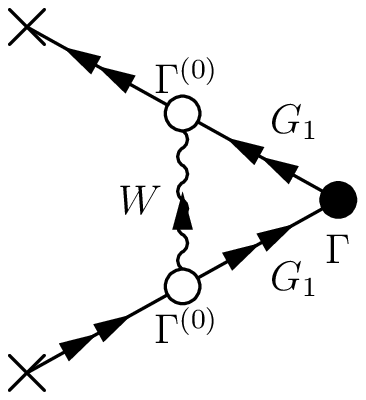,width=80pt}}~.
	\label{eqn:vertex_bse}
\end{equation}
Solving Eq.~(\ref{eqn:dse_full}) and Eq.~(\ref{eqn:vertex_bse})
simultaneously is a formidable task. 
Considerable simplification  of the
 problem is obtained by
replacing the full vertex $\Gamma$ in Eq.~(\ref{eqn:dse_full}) by the
bare vertex $\Gamma^{(0)}$. This is the so-called 
 $GW$ approximation,
\begin{equation}
	\Sigma^{GW}(\vec p,z_\nu)=
	\parbox{80pt}{\psfig{figure=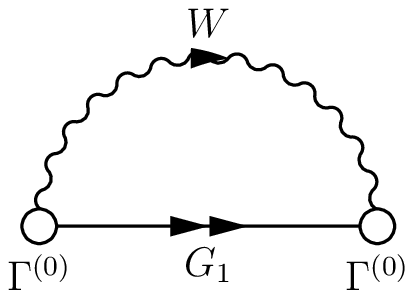,width=80pt}}~.
	\label{eqn:dse}
\end{equation}
The $GW$ approximation is well studied in condensed matter physics. It
also
describes bremsstrahlung an can be further improved
accounting for plasma effects, see Ref.~\cite{Fortmann_CPP_accepted_2006}.

The two-particle contribution $\Pi_2(\vec k, \omega)$ to
the polarization function is described in analogy to
Eq.~(\ref{eqn:pi1_diagram}),
\begin{equation}
	\Pi_2(\vec k,\omega_\mu)=\parbox{100pt}{%
	\includegraphics[width=100pt]{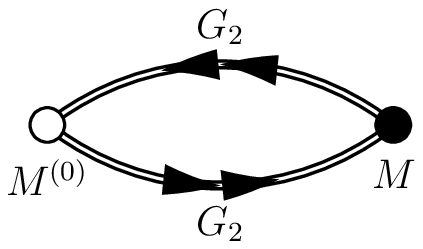}}~,
	\label{eqn:Pi_2_diagram}
\end{equation}
where we have the full two-particle propagator 
\begin{align}
	\nonumber
	G_2(\alpha\vec P,\Omega_\lambda)&=
	\parbox{80pt}{%
	\psfig{figure=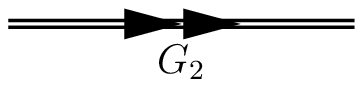,width=80pt}}
	=\parbox{80pt}{%
	\psfig{figure=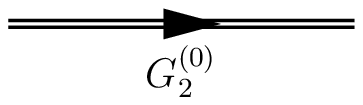,width=80pt}}
	+\parbox{100pt}{%
	\psfig{figure=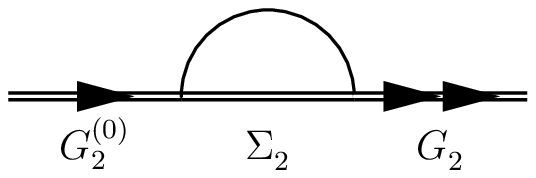,width=100pt}}\\
	&=\frac{1}
	{\Omega_\lambda-E_{\alpha\vec P}-\Sigma_2(\alpha\vec
	P,\Omega_\lambda)}, 
	\label{eqn:G_2_diagram}
\end{align}
with the two-particle self-energy $\Sigma_2(\alpha\vec P,\Omega_\lambda)$,
\begin{equation}
	\Sigma_2(\alpha\vec P,\Omega_\lambda)=\parbox{80pt}{%
	\psfig{figure=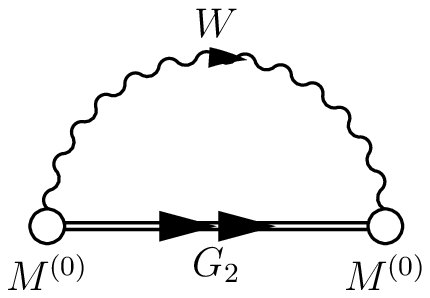,width=80pt}}
	\label{eqn:sigma_2_gw}
\end{equation}
and the vertex follows from an equation in analogy to
Eq.~(\ref{eqn:vertex_bse})
\begin{equation}
	\parbox{50pt}{%
	\psfig{figure=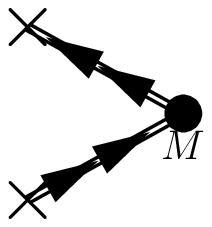 ,width=50pt}}=
	\parbox{50pt}{%
	\psfig{figure=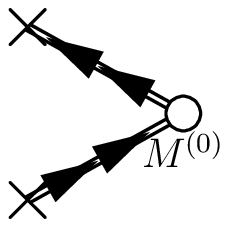 ,width=50pt}}+
	\parbox{100pt}{%
	\psfig{figure=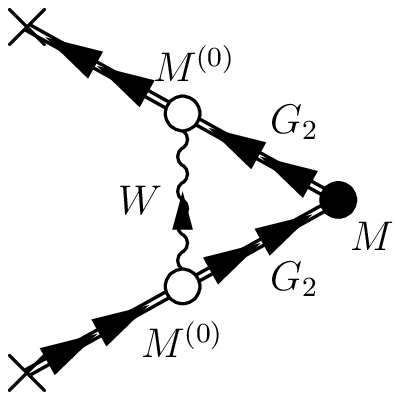 ,width=100pt}}~.
	\label{eqn:bse-twoparticle}
\end{equation}
For the polarization function which defines the screened interaction
used in the two-particle self-energy Eq.~(\ref{eqn:sigma_2_gw}), we
perform the cluster decomposition as outlined in
Sec.~\ref{sec:theory_lineshapes}. In the impact-approximation, we
replace the full screened interaction by the first non-ideal term in
the iteration of Eq.~(\ref{eqn:Vscreened_diagrams}). Diagrammatically,
the two-particle self-energy in impact-approximation reads
\begin{equation}
	\Sigma^\mathrm{imp}(\alpha\vec P,\Omega_\lambda)=
	\parbox{80pt}{%
	\psfig{figure=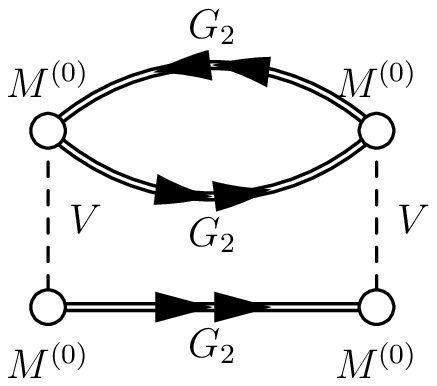 ,width=100pt}}
	\label{eqn:sigma_impact_diagram}
\end{equation}
Note that
one has to take into account that double counting has to be avoided so
that scattering states in $\Pi_2(\vec k, \omega)$ should not interfer
with contributions from $\Pi_1(\vec k, \omega)$. Important are the
account of bound states in $\Pi_2(\vec k, \omega)$. 
 
The higher order corrections to the full vertex $M(\vec q)$ can be
shown to effectively reduce the two-particle
self-energy \cite{guen:habil95} by the amount $i\Gamma^v$. After analytic
continuation $\omega_\mu\to\omega+i0^+$ to real frequencies, we obtain
\begin{multline}
	\Pi_2(\vec k,\omega)=4\sum_{\alpha_1\alpha_2,\vec
	P}\left|M_{\alpha_1\alpha_2}^{(0)}(\vec
	k)\right|^2
	\left(g(E^{(0)}_{\alpha_1\vec P})-g(E^{(0)}_{\alpha_2\vec P-\vec
            k})\right)
	    \\\times 
	    \bigg[\omega-E^{(0)}_{\alpha_2\vec P-\vec k}+E^{(0)}_{\alpha_1\vec
          P}-\mathrm{Re}\,\left\{ 
	  \Sigma_{\alpha_2}(E^{(0)}_{\alpha_1\vec
	  P}+\omega)-\Sigma_{\alpha_1}(E^{(0)}_{\alpha_1\vec P})
	\right\}
	+i\mathrm{Im}\,\left\{\Sigma_{\alpha_2}(E^{(0)}_{\alpha_1\vec
	P}+\omega)+\Sigma_{\alpha_1}(E^{(0)}_{\alpha_1\vec P})
	\right\} +i\Gamma^v\bigg]^{-1}~.
	\label{eqn:Pi_2_sigma}
\end{multline}
Note, that the dependance on $\omega$ is neglected in the self-energy
of the lower energy level \cite{guen:habil95}.
Neglecting the self-energy and the effective vertex in
Eq.~(\ref{eqn:Pi_2_sigma}), we arrive at Eq.~(\ref{eqn:Pi_20_sigma}).

\section{Bound state contribution to the two-particle
self-energy\label{app:bound}}
The interaction between two atoms in state
$\left|\alpha_1\!\right.\rangle$ and $\left|\alpha_2\!\right.\rangle$,
moving with momentum $\vec P_1$ and $\vec P_2$ respectively, and
carrying Matsubara frequencies $\Omega_\lambda$ and $\Omega_\kappa$, upon
exchange of momentum $\vec k$
is given by the following diagram:
 \begin{gather}
	 V_{\alpha_1,\alpha_2}(\vec P_1,\Omega_\lambda,\vec
 	P_2,\Omega_\kappa,\vec k,\omega_\mu)
 	=\parbox{100pt}{\psfig{width=100pt,figure=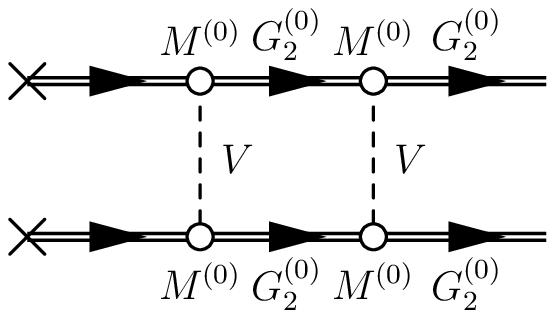}}
 	\\
 	=\sum_{  \alpha_3,\alpha_4 \vec k'} 
	M^{(0)}_{\alpha_1\alpha_3}(\vec k')M^{(0)}_{\alpha_3\alpha_1}(\vec k-\vec k')
	M^{(0)}_{\alpha_2\alpha_4}(-\vec k')M^{(0)}_{\alpha_4\alpha_2}(\vec k+\vec k')
 	V_{\vec k'}V_{\vec k-\vec k'}
	\left[ \Omega_\lambda+\Omega_\kappa-E^{(0)}_{\alpha_3\vec P_1+\vec
	k'}-E^{(0)}_{\alpha_4\vec P_2-\vec k'} \right]^{-1}~,
 	\label{eqn:V_vdw_diagram}
 \end{gather}
 with the unscreened Coulomb propagator $V_{\vec k}=-1/\epsilon_0k^2\Omega_0$.
Due to the ion's heavy mass, we neglect the transfer momentum
$\vec k$ in the intermediate propagator as well as the kinetic energy
$\hbar^2P^2/2M$. Replacing the Matsubara frequencies $\Omega_\lambda$
and $\Omega_\kappa$ by their on-shell values, we can perform the
summations over momenta in Eq.~(\ref{eqn:matrixelementM_p}) and arrive
at the familiar expression Eq.~(\ref{eqn:V_vdw_R}).
\section{Calculation of the interaction
strength\label{app:interaction_strength}}
The expression Eq.~(\ref{eqn:V_vdw_R_dip}) has to be evaluated. We
rewrite it as
\begin{align}
	\begin{split}
		V^\mathrm{vdW}_{\alpha_1,\alpha_2}(R)&=
		-\frac{C_6(\alpha_1,\alpha_2)}{R^6}\\
		&=-\frac{e^4}{(4\pi\epsilon_0)^2R^6}\sum_{\alpha_3\alpha_4}{}^{'}\frac{\left|\langle
		\alpha_1\alpha_2\left|\mathrm{W}\right|\alpha_3\alpha_4\rangle\right|^2}{E^{(0)}_{\alpha_1}+E^{(0)}_{\alpha_2}-E^{(0)}_{\alpha_3}-E^{(0)}_{\alpha_4}}~.
	\end{split}
	\label{eqn:deriv_strength010}
\end{align}
The primed sum indicates summation only over
states, which assure a finite denominator.

Henceforth, we assume one atom in an excited state 
$\alpha_1=n_1l_1, n_1=2,3,4,\dots$, and the second atom in its ground state, 
$\alpha_2\equiv 0$. The denominator is dominated by the term 
$E^{(0)}_{\alpha_1}+E^{(0)}_{\alpha_2}=-\mathrm{Ry}(1+1/n_1^2)$. 
Eq.~(\ref{eqn:deriv_strength010}) becomes
\begin{align}
	C_6(\alpha_1,0)&=\left(\frac{e^2}{4\pi\epsilon_0}\right)^2
	\frac{1}{\left( 1+1/n_1^2 \right)\mathrm{Ry}}
	\sum_{\alpha_3\alpha_4}{}^{'}
	\left|\langle\alpha_10\left|\mathrm{W}\right|\alpha_3\alpha_4\rangle\right|^2\\
	&=\left(\frac{e^2}{4\pi\epsilon_0}\right)^2
	\frac{1}{\left( 1+1/n_1^2 \right)\mathrm{Ry}}
	\sum_{\alpha_3\alpha_4}
	\langle \alpha_10\left|\mathrm{W}\right|\alpha_3\alpha_4\rangle
	\langle\alpha_3\alpha_4\left|\mathrm{W}\right|\alpha_10 \rangle-
	\left|\langle\alpha_1 0\left|\mathrm{W}\right|\alpha_10 \rangle\right|^2\\
	&=\left(\frac{e^2}{4\pi\epsilon_0}\right)^2
	\frac{1}{\left( 1+1/n_1^2 \right)\mathrm{Ry}}
	\langle \alpha_10\left|\mathrm{W}^2\right|\alpha_10 \rangle~.
	\label{eqn:deriv_strength020}
\end{align}
The diagonal matrix element of $\mathrm{W}$ vanishes for 
atoms with no permanent electric dipole moment.
Chosing the nucleus-nucleus axis parallel to the $x$-axis of the
coordinate system, the operator $\mathrm{W}$ reads
\begin{equation}
	\mathrm{W}=\mathrm{y}_1\mathrm{y}_2+\mathrm{z_1}\mathrm{z}_2-2\mathrm{x}_1\mathrm{x}_2~.
	\label{eqn:W_op}
\end{equation}
Since the expectation values of any coordinate $x,y$ or $z$ vanishes,
e.g. $\langle\alpha_10\left|\mathrm{r_i}\right|\alpha_10\rangle=0,\,i=1,2,3$,
the mixed terms in $\mathrm{W}^2$ vanish. We obtain
\begin{align}
	C_6(\alpha_1,0)&=\left(\frac{e^2}{4\pi\epsilon_0}\right)^2
	\frac{1}{\left( 1+1/n_1^2 \right)\mathrm{Ry}}
	\langle\alpha_10\left|\mathrm{y_1^2y_2^2+z_1^2z_2^2+4x_1^2x_2^2}\right|\alpha_10\rangle\\
	&=\left(\frac{e^2}{4\pi\epsilon_0}\right)^2
	\frac{1}{\left( 1+1/n_1^2 \right)\mathrm{Ry}}\Big[\\
	&\langle0\left|\mathrm{y_1^2}\right|0\rangle\langle
	\alpha_1\left|\mathrm{y_2^2}\right|\alpha_1\rangle+\langle0\left|\mathrm{z_1^2}\right|0\rangle\langle
	\alpha_1\left|\mathrm{z_2^2}\right|\alpha_1\rangle+4\langle0\left|\mathrm{x_1^2}\right|0\rangle\langle
	\alpha_1\left|\mathrm{x_2^2}\right|\alpha_1\rangle\Big]~,\\
	\intertext{and with}
	\langle r_i^2\rangle=\frac{1}{3}\langle r^2\rangle\\
	C_6(\alpha_1,0)&=\left(\frac{e^2}{4\pi\epsilon_0}\right)^2
	\frac{1}{\left( 1+1/n_1^2 \right)\mathrm{Ry}}
	\frac{2}{3}\langle0\!\left|\,\mathrm{r}^2\,\right|\!0\rangle\langle
	\alpha_1\!\left|\,\mathrm{r^2}\,\right|\!\alpha_1\rangle~.
	\label{eqn:deriv_strength030}
\end{align}
Using
\begin{equation}
	\langle
	nl\!\left|\,\mathrm{r^2}\,\right|\!nl\rangle=\frac{n^2
	a_\mathrm{B}^2}{2}\left[ 5n^2+1-3l(l+1) \right]~,
	\label{eqn:r2}
\end{equation}
we finally obtain Eq.~(\ref{eqn:int_strength}).

Note, that for the case, that the first atom is also in its ground
state, i.e. $\alpha\equiv 0$, $C_6(0,0)=12$ is obtained. The difference to
the London-Eisenschitz result 12.94 is due to the approximative
treatment of the denominator in Eq.~(\ref{eqn:deriv_strength010})
 in the calculation presented above. For
excited states, which are considered in this work,  
this contribution  becomes negligible.

\section{Evaluation of the van der Waals profile\label{app:vdw-profile}}
We start with the general form of 
expression (\ref{eqn:V_vdw_R}), i.e. the intensity
distribution of a given spectral line, due to interaction
between the radiator, which is in state
$\left|\,i\,\right.\rangle$ before the transition and in state
$\left|\,f\,\right.\rangle$ after the transition and the
perturber $i, i=1\dots N$ in state
$\left|n_i\!\right.\rangle$ .
It is obtained by
performing an averaging
procedure over the distribution of the $N$ perturbing atoms,
\begin{equation}
	I_{if}(\Delta\omega)=
	\int\mathrm{d}^3\vec r_1\int\mathrm{d}^3\vec
	r_2\ldots\int\mathrm{d}\vec r_N P_N(\vec r_1,\vec
	r_2,\ldots,\vec r_N)\,\delta\left( \sum_{j=1}^N
	V_{i\alpha_j}(r_j)-V_{f\alpha_j}(r_j)-\Delta\omega \right)~.
	\label{eqn:app:profile_at-at}
\end{equation}
Here, $P_N(\vec r_1,\ldots\vec r_N)$ is the probability density for
having atom 1 in the volume element $\mathrm{d}^3\vec r_1$ at $\vec
r_1$, atom 2 in the volume element $\mathrm{d}^3\vec r_2$ at $\vec
r_2$, and so forth. In can be expanded in a cluster decomposition as 
\begin{equation}
	P_N(\vec r_1,\vec r_2,\ldots,\vec r_N)=
	\frac{1}{V^N}\left( 1+\sum_{j<k}g_2(\vec r_j,\vec
	r_k)+\sum_{j<k<l}g_3(\vec r_j,\vec r_k, \vec r_l)+\dots
	\right)~,
\end{equation}
where the lowest non-ideal term is the pair distribution function,
giving the probability to find a second particle at $\vec r_k$ if there
is one particle at $\vec r_j$.

The $N^3$-fold integral in Eq.~(\ref{eqn:app:profile_at-at}) can be 
evaluated using the Fourier
representation of the delta function,
\begin{equation}
	\delta(\Phi-\Delta\omega)=\frac{1}{2\pi}\int_{-\infty}^\infty
	\mathrm{d}\rho\,\mathrm{e}^{i\rho(\Phi-\Delta\omega)}~,
	\label{eqn:delta_fourier}
\end{equation}
with
$\Phi=\sum_{j=1}^N [V_{i\alpha_j}(r_j)-V_{f\alpha_j}(r_j)]$.
In the case of statistically independant atoms, 
which are all in the
same quantum state $\alpha_j\equiv \alpha_p$ i.e. $P_N(\vec
r_1,\vec r_2,\ldots,\vec r_N)=V^{-N}$, one immediatly finds
\begin{equation}
	I_{if}(\Delta\omega)=\frac{1}{2\pi}\left(
	\frac{4\pi}{V} \right)^N
	\int_{-\infty}^\infty\mathrm{d}\rho\,\mathrm{e}^{-i\Delta\omega\rho}
	\left[ \int\mathrm{d}r\,r^2\mathrm{e}^{i\rho
	[V_{i,\alpha_p}(r)-V_{f,\alpha_p}(r)]} \right]^{N}~.
	\label{eqn:profile_random}
\end{equation}
Using the van der Waals limit of the atom-atom potential, Eq.~(\ref{eqn:vdwpot_00}),
Eq.~(\ref{eqn:profile_random}) turns to 
\begin{equation}
	I^\mathrm{vdW}_{if}(\Delta\omega)=\frac{1}{2\pi}\left(
	\frac{4\pi}{V} \right)^N
	\int_{-\infty}^\infty\mathrm{d}\rho\,\mathrm{e}^{-i\Delta\omega\rho}
	\left[ \int\mathrm{d}r\,r^2\mathrm{e}^{-i\rho C_6/r^6
	} \right]^{N}~.
	\label{eqn:profile_random_dipole}
\end{equation}
The integration over $r$ can be rewritten in the form
\begin{align}
	\int_0^\infty\mathrm{d}r\,r^2\,\mathrm{e}^{-i\rho C_6/r^6}&=
	\int_0^{\infty}\mathrm{d}r\,r^2\left[ 1-\left(
	1-\mathrm{e}^{-i\rho C_6/r^6} \right) \right]\\
	&=\frac{V}{4\pi}\left( 1-\frac{4\pi\tilde V(\rho)}{V} \right)~,
	\label{eqn:app:profile_integration_1}
\end{align}
with $C_6=C_6(f,0)-C_6(i,0)$.
In the thermodynamic limit, i.e. $V\to\infty,
N\to\infty$, while keeping constant the density of neutrals
$N/V=n_\mathrm{atom}=\mathrm{const}$, one obtains 
\begin{equation}
	\lim_{N\to\infty}\left( 1-\frac{4\pi\tilde V(\rho) n_\mathrm{atom}}{N} \right)^N=
	\mathrm{e}^{-4\pi n_\mathrm{atom} \tilde V(\rho)}~.
	\label{eqn:app:thermodynamic_limit}
\end{equation}
For $\tilde V(\rho)$,
\begin{equation}
	\tilde V(\rho)=\int_0^{\infty}\mathrm{d}r\,r^2\,\left(
	1-\mathrm{e}^{-i\rho C_6/r^6} \right)=\frac{(2\pi
	C_6\rho)^{1/2}}{6}(1-i)~,
	\label{eqn:app:integration_r}
\end{equation}
is obtained, and finally the Fourier integral
Eq.~(\ref{eqn:profile_random_dipole}) can be evaluated to give
\begin{equation}
	I^\mathrm{vdW}_{if}(\Delta\omega)=\frac{2\pi n_\mathrm{atom} \sqrt{C_6}}{3
	(-\Delta\omega)^{3/2}}\mathrm{e}^{\pi C_6 (2\pi
	n_\mathrm{atom}/3)^2/\Delta\omega}~,
	\label{eqn:app:margenau_final}
\end{equation}
which corresponds to Eq.~(\ref{eqn:profile_london}).

\end{appendix}
\end{document}